\documentclass[nofootinbib,aps,prl,twocolumn,preprintnumbers,amsmath,amssymb,superscriptaddress,10pt,floatfix]{revtex4-2}

\usepackage{graphicx}
\usepackage{bm}
\usepackage{amsmath}
\usepackage{braket}
\usepackage{placeins}
\usepackage{multirow}
\usepackage{makecell}
\usepackage{soul}
\usepackage{braket}
\usepackage[shortlabels]{enumitem}
\usepackage{placeins}
\usepackage[hidelinks]{hyperref}
\usepackage{xcolor}
\usepackage{orcidlink}

\usepackage[labelfont=bf, textfont=it, font=small]{caption}
\newcommand{\bee}{\begin{equation}}
\newcommand{\eee}{\end{equation}}
\newcommand{\beq}{\begin{eqnarray}}
\newcommand{\eeq}{\end{eqnarray}}
\newcommand{\beqnn}{\begin{eqnarray*}}
\newcommand{\eeqnn}{\end{eqnarray*}}
\newcommand{\vs}[1]{\vspace{#1\baselineskip}}

\renewcommand{\thesection}{\arabic{section}}

\newcommand{\sst}[1]{{\scriptscriptstyle{\mathrm{#1}}}}
\newcommand{\Tr}{\ensuremath{\mathrm{Tr}}}
\newcommand{\chip}{\chi^\prime}
\newcommand{\etap}{\eta^\prime}
\newcommand{\gA}{g^{\sst{(0)}}_{\sst{A}}}
\newcommand{\Nf}{N_{\sst{f}}}
\newcommand{\Rs}{R_{\sst{s}}}
\newcommand{\YM}{{\sst{YM}}}
\newcommand{\cool}{{\sst{cool}}}
\newcommand{\SU}{\mathrm{SU}}
\newcommand{\QCD}{{\sst{QCD}}}

\renewcommand{\L}{{\sst{L}}}
\newcommand{\LO}{{\sst{LO}}}
\newcommand{\rep}{{\sst{r}}}
\newcommand{\defect}{{\sst{d}}}
\newcommand{\NN}{\mathrm{N}}
\newcommand{\dd}{\mathrm{d}}
\newcommand{\ee}{\mathrm{e}}
\newcommand{\ii}{\mathrm{i}}

\newcommand{\sub}{{\sst{(sub)}}}
\newcommand{\one}{{\sst{(1)}}}
\newcommand{\chipsub}{\chi^{\prime\,\sub}_{_\YM}}

\graphicspath{{figs/}}

\begin{document}
	
\title{The topological susceptibility slope $\chi^\prime$ in the large-$N$ limit}

\author{Claudio Bonanno\,\orcidlink{0000-0002-1421-6510}}
\email{claudio.bonanno@unibe.ch}
\email{\\}
\affiliation{\footnotesize Albert Einstein Center for Fundamental Physics, Institute for Theoretical Physics, University of Bern, Sidlerstra{\ss}e 5, CH-3012 Bern, Switzerland}
\affiliation{\footnotesize Instituto de F\'isica Te\'orica UAM-CSIC, c/ Nicol\'as Cabrera 13-15, Universidad Aut\'onoma de Madrid, Cantoblanco, E-28049 Madrid, Spain}

\begin{abstract}
{\footnotesize
This paper presents the first non-perturbative lattice determination of the Yang--Mills topological susceptibility slope $\chi^\prime$ in the large-$N$ limit. This quantity represents the $\mathcal{O}(p^2)$ term of the momentum expansion of the topological charge density two-point correlator, and has important theoretical and phenomenological implications for strong interactions. This calculation is based on a novel algorithm that avoids topological freezing at large $N$ on fine lattices, and on a novel method to reliably compute $\chi^\prime$ on the lattice. The results of this study are relevant for the description of the proton spin in deep inelastic scattering experiments via the Shore--Veneziano formula.
}
\end{abstract}

\maketitle

\section*{Introduction}

The topological susceptibility slope $\chip$, the second moment of the Euclidean two-point function of the topological charge density $q(x)=\frac{1}{64\pi^2} \varepsilon_{\mu\nu\rho\sigma}G^a_{\mu\nu}(x) G^a_{\rho \sigma}(x)$,
\beq
\chip = \frac{1}{8} \int \dd^4 x \, x^2 \, \braket{q(x)q(0)},
\eeq
is a quantity of great theoretical interest in $\SU(N)$ Yang--Mills (YM) theories ($N$ = number of colors) and in Quantum Chromo-Dynamics (QCD), with fundamental phenomenological implications for strong interactions.

In particular, it is extremely relevant to clarify the origin of the proton spin --- the so-called \emph{proton spin puzzle} --- one of the most important problems in particle and nuclear physics under active investigation~\cite{Bass:2009dr,Aidala:2012mv,Gao:2015aax,NatRevPhysEditorialProtonPuzzle:2021xxx,Liu:2021lke,Zahed:2026wag}.
Deep inelastic scattering measurements of the flavor-singlet axial charge $\gA$,
\beq
S^\mu\gA = \frac{1}{2m_{\sst{N}}} \bra{\NN(P,S)} J_{\sst{5}}^\mu \ket{\NN(P,S)},
\eeq
revealed that this form factor is much smaller than the value predicted by the quark constituent model~\cite{EuropeanMuon:1989yki,HERMES:2006jyl,COMPASS:2006mhr}, where $\gA$ is interpreted as the fraction of the proton spin carried by the quarks (here $J_{\sst{5}}^\mu= \sum_{\sst{f}}\overline{\psi}_{\sst{f}}\gamma_{\sst{5}}\gamma^\mu\psi_{\sst{f}}$, while $P, S^\mu, m_{\sst{N}}$ are the nucleon momentum, spin polarization, and mass). Shore and Veneziano, starting from the anomalous mixing between $\partial_\mu J_{\sst{5}}^\mu(x)$ and $q(x)$, advocated the possibility that the smallness of $\gA$ could be due to the screening of the topological charge in the QCD vacuum, i.e., to the smallness of $\chip$~\cite{Shore:1990zu,Shore:1992xxx,Shore:1997tq} (see also~\cite{Tarasov:2020cwl,Tarasov:2021yll,Tarasov:2025mvn} for a recent derivation using the worldline formalism):
\beq\label{eq:shore_veneziano_formula}
\gA = \frac{\Nf}{m_{\sst{N}}} g_{\etap \sst{N}\sst{N}} \lim_{m \, \to \, 0} \sqrt{\vert \chip \vert},
\eeq
with $g_{\etap \sst{N}\sst{N}}$ an effective $\etap$-nucleon coupling. Equation~\eqref{eq:shore_veneziano_formula}, valid in the chiral limit with $\Nf$ flavors, holds in the so-called OZI limit, which is theoretically justifiable within the large-$N$ limit framework, when the $\etap$ becomes a would-be Goldstone boson and does not mix with gluonia, see~\cite{Shore:1992xxx,Shore:2007yn,Tarasov:2025mvn} and references therein on this point.

Despite its key phenomenological importance, the current status regarding the theoretical determination of $\chip$ is not completely satisfactory. Estimates from QCD Sum Rule~\cite{Ioffe:1998sa,Narison:1998aq,Narison:2006ws,Narison:2021svo}, Chiral Perturbation Theory ($\chi$PT)~\cite{Leutwyler:2000jg}, and the NJL model~\cite{Fukushima:2001ut} are of the same order of magnitude but show some tension among them, see Ref.~\cite{Bonanno:2023ple} for a discussion. Given the non-perturbative nature of topological properties of gauge theories, the lattice is a natural framework to address the determination of $\chip$ from first principles, but so far only a few preliminary and inconclusive investigations have appeared~\cite{DiGiacomo:1990ij,Briganti:1991pb,DiGiacomo:1992wg,Boyd:1997nt}.

The goal of this article is to present the first non-perturbative determination of the topological susceptibility slope in the large-$N$ limit from numerical Monte Carlo simulations
on the lattice, and to discuss the phenomenological implications of this result in relation with the proton spin problem and the Shore--Veneziano formula. This calculation relies on the novel lattice method to compute $\chip$ introduced in $2d$ $\mathrm{CP}^{N-1}$ models in~\cite{Bonanno:2022hmz} (where analytical results up to NLO in the $1/N$ expansion have been reproduced), whose feasibility in $4d$ gauge theories was established in $\SU(3)$ Yang--Mills in~\cite{Bonanno:2023ple}.

\vs{-1.5}
\section*{$\chip$ at large $N$}
\vs{-0.5}

Starting from the usual large-$N$ arguments leading to the Witten--Veneziano equation in the chiral limit,
\bee\label{eq:witten_veneziano}
m_{\etap}^2 = \frac{2 \Nf}{F^2} \chi_{_\YM}, \quad \chi_{_\YM}=\int \dd^4x \, \braket{q(x)q(0)}_{_\YM},
\eee
with $F$ the pion decay constant in the chiral limit (in the ``92 MeV'' convention) and $\chi_{_\YM}$ the Yang--Mills topological susceptibility, the following large-$N$ relation between the QCD and the Yang--Mills susceptibility slopes can be derived in the chiral limit with $\Nf$ flavors~\cite{Giusti:2001xh,Bonanno:2023ple}:
\beq\label{eq:chipQCD_chipYM_formula}
\lim_{m \,\to\, 0} \chip_{_\QCD} = \chip_{_\YM} - \frac{1}{2\Nf}F^2.
\eeq
These equations descend from a large-$N$ expansion of the two-point correlator $C(p^2) = \int \dd^4x \, \ee^{\ii p \cdot x} \braket{q(x)q(0)}$,
\beq\label{eq:largeN_exp_corr}
C_{_\QCD}(p^2) = C_{_\YM}(p^2) - \frac{\vert A_{\etap} \vert^2}{p^2 + m_{\etap}^2} + \mathcal{O}\left(\frac{1}{N}\right),
\eeq
where $A_{\etap}=\bra{0} q(0) \ket{\etap}=\frac{1}{\sqrt{2\Nf}}F m^2_{\etap}$, and where the subscripts ``QCD'' and ``YM'' stand for expectation values taken in full QCD and in pure Yang--Mills theories. The rationale behind Eq.~\eqref{eq:largeN_exp_corr} is that, at leading order in the $1/N$ expansion, one has only the pure-glue and the $\etap$ contributions. Using that:\footnote{For a discussion about the sign conventions in the definitions of $\chi$ and $\chip$, see the Supplemental Material~\cite{suppmat}.}
\beq
\chi = C(p^2=0), \qquad \chip = - \frac{\dd C(p^2)}{\dd p^2}\bigg\vert_{p^2\,=\,0},
\eeq
Eqs.~\eqref{eq:witten_veneziano} and~\eqref{eq:chipQCD_chipYM_formula} are obtained from Eq.~\eqref{eq:largeN_exp_corr} in the chiral limit (where $\chi_{_\QCD}=0$).

The large-$N$ counting of Eq.~\eqref{eq:shore_veneziano_formula} reveals that
\beq
\chip_{_\QCD}\sim\mathcal{O}(N)
\eeq
since $m_{\sst{N}}\sim\mathcal{O}(N)$, $\gA\sim\mathcal{O}(N^0)$~\cite{Hidaka:2010ph,Kojo:2012hf,Chen:2017tgv,Richardson:2022hyj}, $g_{\etap\sst{N}\sst{N}}\sim\mathcal{O}(\sqrt{N})$~\cite{Shore:2006mm} (the same large-$N$ scaling for $\chip_{_\QCD}$ can be predicted from a chiral effective theory approach~\cite{Leutwyler:2000jg}). Concerning the right-hand side of Eq.~\eqref{eq:chipQCD_chipYM_formula}, while $F^2 \sim \mathcal{O}(N)$~\cite{Witten:1979vv,Veneziano:1979ec}, it was argued in~\cite{Giusti:2001xh} that, just like $\chi_{_\YM}$,
\beq
\chip_{_\YM}\sim \mathcal{O}(N^0)
\eeq
for consistency with the Witten--Veneziano argument. This would be similar to what happens in $2d$ $\mathrm{CP}^{N-1}$ models, where $\chi$ and $\chip$ have the same large-$N$ scaling~\cite{Campostrini:1991kv,Campostrini:1991tw}. If this statement is correct, it implies that gluons only give a non-vanishing contribution to $\chip_{_\QCD}$ via Eq.~\eqref{eq:chipQCD_chipYM_formula} at large but finite $N$, while in the strict large-$N$ limit their contribution vanishes, and only the $\etap$ one survives. If this scenario is realized, then $\chip_{_\QCD}$ at $N=\infty$ is fully determined by the large-$N$ limit of $F$.

The main objective of this paper is to provide the first lattice calculation of the large-$N$ limit of $\chip_{_\YM}$ and investigate its large-$N$ behavior. The final goal is to combine the outcome of this calculation with the recent lattice large-$N$ result for $F$~\cite{Bonanno:2025hzr} obtained via the twisted Eguchi--Kawai formulation to finally obtain $\chip_{_\QCD}$ at large $N$ from Eq.~\eqref{eq:chipQCD_chipYM_formula}. Since the latter is exactly the quantity entering the Shore--Veneziano formula~\eqref{eq:shore_veneziano_formula}, this will allow a direct comparison of this equation with experimental data.

\vs{-1}
\section*{Lattice methods}
\vs{-0.5}

To compute $\chip_{_\YM}$ at large $N$, simulations were performed for $N=3,4,5,6$ using the standard plaquette Yang--Mills action on a hypercubic periodic lattice, keeping a constant lattice volume $\ell^4\simeq(\text{1.5 fm})^4$ in all cases (the conversion from lattice to physical units of dimensionful quantities will be done via the string tension $\sigma$). For all values of $N$, 4--5 values of the lattice spacing $a$ were considered to enable a continuum limit extrapolation. Simulated values of the lattice spacing were drawn from a uniform range across all $N$ values, $0.1 \text{ fm} \gtrsim a \gtrsim 0.05 \text{ fm}$, to ensure that lattice artifacts are under control as the large-$N$ limit is approached. This choice would be computationally prohibitive when adopting standard local Monte Carlo algorithms due to the infamous \emph{topological freezing} problem~\cite{Alles:1996vn,DelDebbio:2002xa,DelDebbio:2004xh,
Schaefer:2010hu}.

When approaching the continuum limit $a\to 0$, the autocorrelation time $\tau_Q$ of the topological charge $Q=\int \dd^4 x \, q(x)$ (i.e., the number of Monte Carlo steps necessary to draw two uncorrelated samples of $Q$) grows rapidly as a function of $1/a$ and of $N$. This makes it unfeasible with standard methods to reliably sample the topological charge distribution at large $N$ even at coarse lattice spacing. While in principle $\chip$, being the second moment of the two-point function $\braket{q(x)q(0)}$, could be computed even in the presence of a fixed topological charge, its value turns out in practice to be strongly correlated with the topological background~\cite{Bonanno:2022hmz,Bonanno:2023ple}. Thus, a sampling problem in $Q$ would lead to a large bias on the actual physical value of $\chip$, manifesting as somewhat large power-like finite-size effects instead of exponentially-suppressed ones~\cite{Brower:2003yx}.

To circumvent this problem, I adopted the \emph{Parallel Tempering on Boundary Conditions} algorithm~\cite{Hasenbusch:2017unr,Bonanno:2020hht,Bonanno:2024zyn}. This algorithm efficiently combines simulations with periodic boundary conditions (PBC) and open boundary conditions (OBC) in a parallel tempering framework, allowing to dramatically mitigate the $N$- and $a$-scaling of $\tau_Q$, and thus to avoid topological freezing in the parameter space explored here. This algorithm was shown in~\cite{Bonanno:2025eeb} to outperform, at fixed computational effort, both simulations done with PBC or OBC. In that study, the algorithm was employed to compute the topological susceptibility $\chi_{_\YM}$ in the large-$N$ limit from finer lattice spacings compared to those employed in previous PBC and OBC studies~\cite{Bonati:2015sqt,Bonati:2016tvi,Athenodorou:2020ani,Athenodorou:2021qvs,Ce:2015qha,Ce:2016awn}. Here, the same simulations of~\cite{Bonanno:2025eeb} are used for the computation of $\chip_{_\YM}$. More technical details are reported in the Supplemental Material~\cite{suppmat}.

The lattice topological susceptibility slope $\chip_{_\YM}$ is computed from the following discretized expression:
\beq
a^2 \chip_{_\YM} = \frac{1}{8} \sum_{x} d^2(x,0)\braket{q_{_\L}(x) q_{_\L}(0)}_{_\YM},
\eeq
with $d(x,0)$ the shortest distance (in the Euclidean metric) between sites $x$ and $0$ on a periodic lattice. The lattice topological charge density $q_{_\L}(x)$ is formulated via the standard \emph{clover} discretization, and it is computed on \emph{smoothened} gauge fields. Smoothing is used to suppress UV fluctuations at the scale of the lattice spacing, which would otherwise lead to finite multiplicative and divergent additive renormalizations of the two-point correlator of $q(x)$~\cite{Campostrini:1989dh,DElia:2003zne,Vicari:2008jw}. Several smoothing methods have been employed in the literature such as cooling~\cite{Berg:1981nw, Iwasaki:1983bv, Teper:1985rb,Ilgenfritz:1985dz, Campostrini:1989dh}, gradient flow~\cite{Narayanan:2006rf,Luscher:2009eq,Luscher:2010iy} or stout smearing~\cite{Morningstar:2003gk}. They have all been shown to yield perfectly agreeing results once the smoothing radius $\Rs$, the length scale below which UV fluctuations are damped, is matched among different methods~\cite{Alles:2000sc,Bruckmann:2006wf,Bonati:2014tqa,Alexandrou:2017hqw}. In the case of Wilson cooling (the choice adopted in this study), the smoothing radius reads $\Rs=a\sqrt{8 n_\cool/3}$ (with $n_\cool$ the number of cooling steps). This can be matched with $\Rs=\sqrt{8 t}$ ($t=$ flow time) for the Wilson gradient flow~\cite{Bonati:2014tqa} (see also~\cite{Alexandrou:2015yba,Alexandrou:2017hqw}).

While $Q$ and observables depending only on $Q$ such as the topological susceptibility $\chi = \braket{Q^2}/V$ are expected~\cite{Ce:2015qha} and have been shown numerically~\cite{Ce:2015qha,Giusti:2018cmp,Bonanno:2025eeb,Durr:2025qtq} to be independent of $\Rs$ once the continuum limit is taken, this is not the case for $\braket{q(x)q(0)}$~\cite{Ce:2015qha,Altenkort:2020axj}. After the continuum limit is taken, $C_{\YM}(p^2)$ and thus $\chip_{_\YM}$ will exhibit a non-trivial dependence on $\Rs$~\cite{Bonanno:2022hmz,Bonanno:2023ple}. Based on the results of~\cite{,Ce:2015qha,Altenkort:2020axj}, a leading linear running of $\chip$ with $\Rs^2$, due to the regular part of $C_{\YM}(p^2)$, is expected. In addition, a \emph{singular} $1/\Rs^2$ term can be predicted, due to the $\mathcal{O}(p^2)$ contact term of $C_{\YM}(p^2)$~\cite{Giusti:2001xh}:
\bee\label{eq:chip_running_general}
\chip_{_\YM}(N,\Rs) = \frac{S(\Rs)}{\Rs^2} + \chip_{_\YM}(N)[1+ \mathcal{O}(\Rs^2)],
\eee
with $\chip_{_\YM}(N)$ the \emph{scale-invariant} susceptibility slope, and $S(\Rs)$ the dimensionless coefficient of the singular term. The following prescription to subtract the singularity and extract the scale-invariant $\chip$ will be used:
\bee\label{eq:chip_subtraction_prescription}
\chip_{_\YM}(N) \equiv \lim_{\Rs \to 0}\lim_{a\to 0} \underbrace{\chip_{_\YM}(N,\Rs,a) - \frac{S_{\LO}(\Rs,a)}{\Rs^2}}_{\text{\normalsize $\chipsub(N,\Rs,a)$}},
\eee
with $S_{\LO}$ the coefficient of the singular term computed at leading order in perturbation theory, and evaluated via a lattice definition of the coupling obtained from the action density after cooling (more details are reported in the Supplemental Material~\cite{suppmat}). Clearly, a thorough stability check of the $\Rs\to0$ extrapolation is necessary to verify that neglected higher-order terms in the perturbative expansion of $S$ are negligible in the employed range of smoothing radii.

\vs{-0.7}
\section*{Results}
\vs{-0.6}

\begin{figure}[!t]
\centering
\includegraphics[scale=0.25]{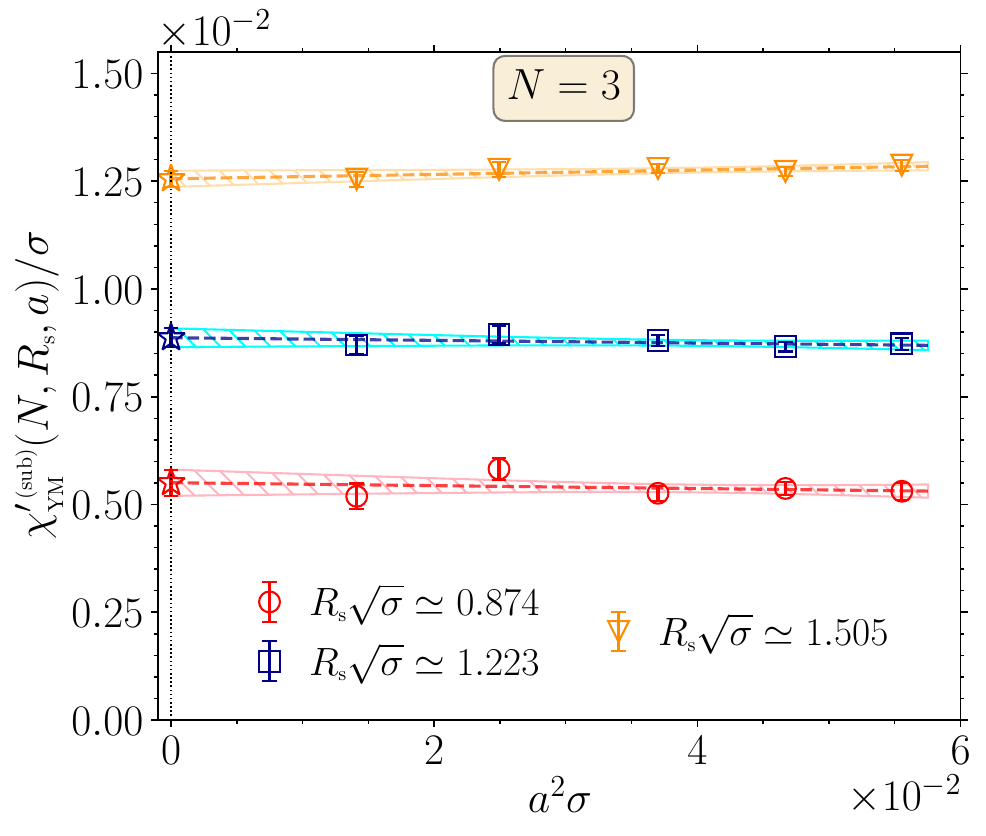}
\includegraphics[scale=0.25]{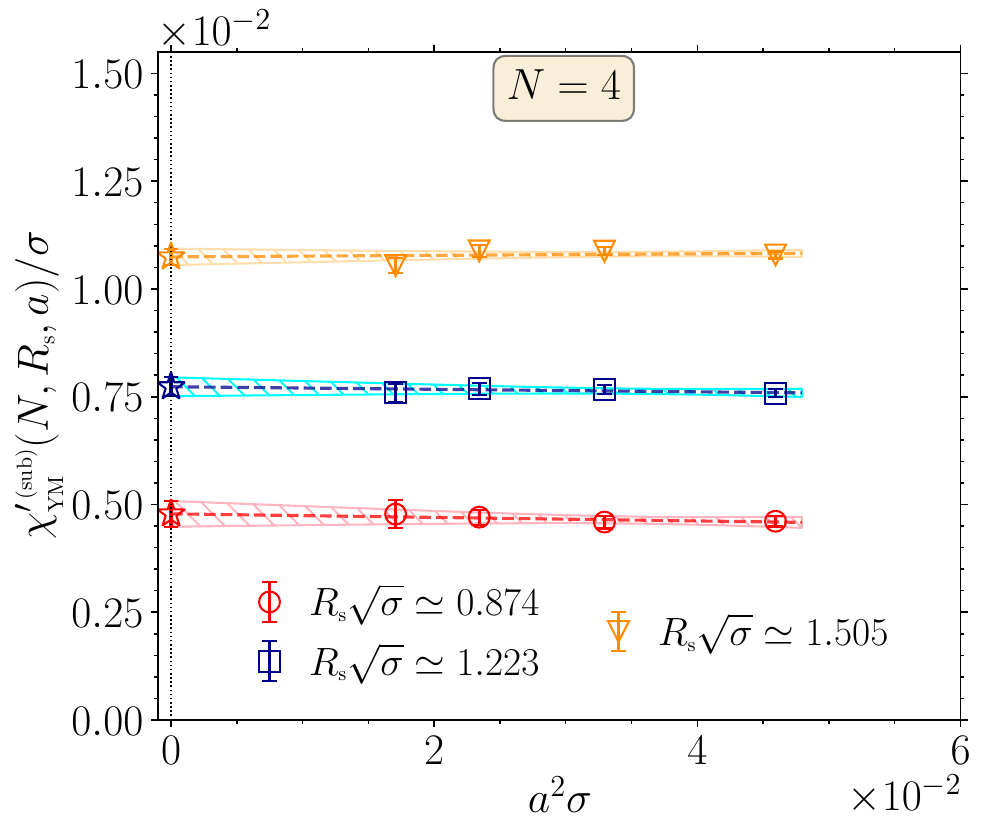}
\includegraphics[scale=0.25]{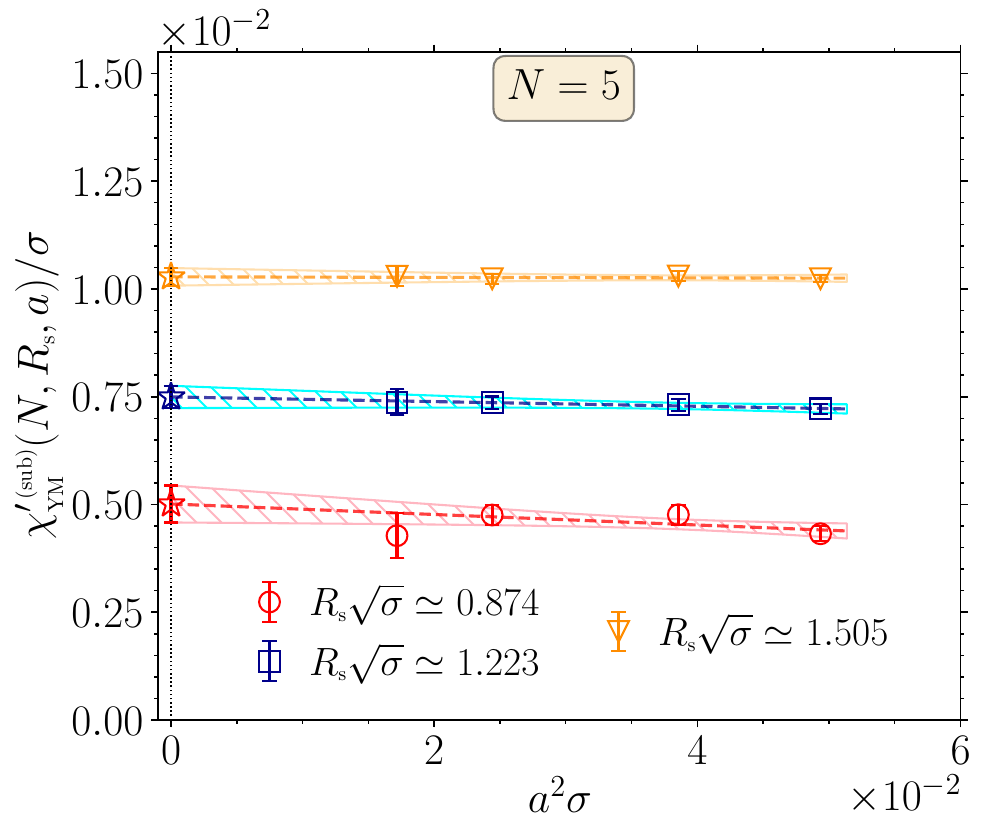}
\includegraphics[scale=0.25]{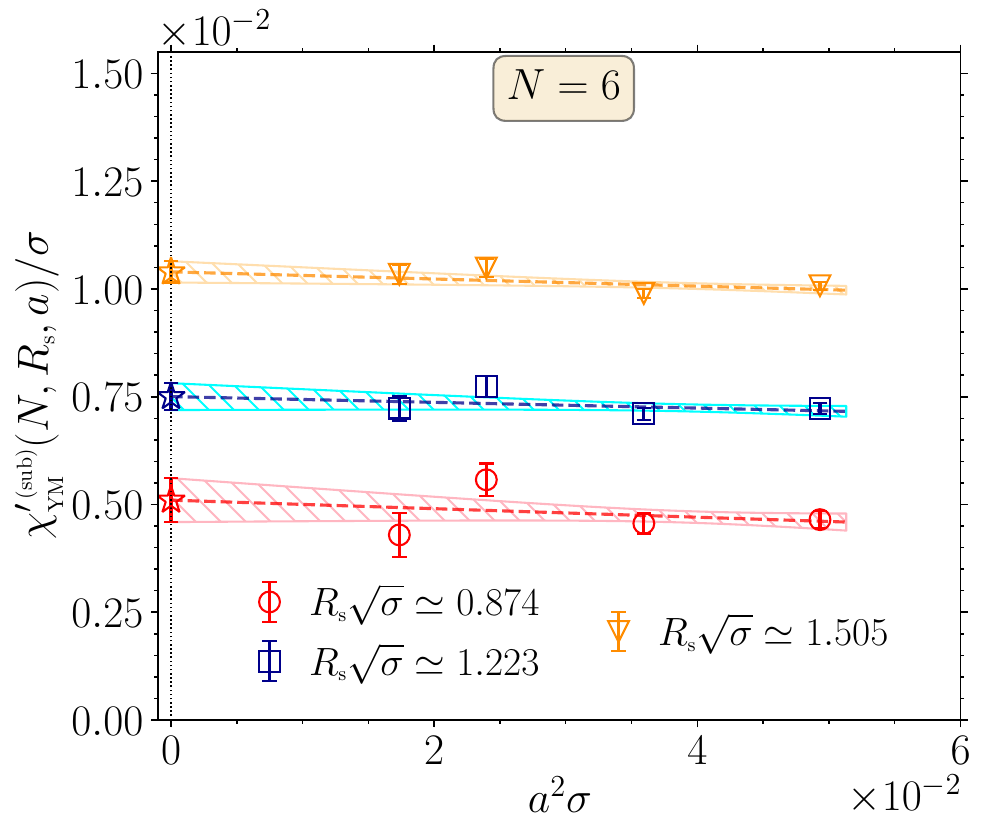}
\caption{Continuum limits of $\chipsub(N,\Rs,a)$ in units of the string tension $\sigma$ for $N=3,4,5,6$ and for three values of the smoothing radius $\Rs\sqrt{\sigma}\simeq 0.874,1.223,1.505$. Star points in $a^2\sigma=0$ represent the continuum extrapolations.}
\label{fig:cont_limit}
\end{figure}

The topological susceptibility slope $\chip_{_\YM}(N,\Rs,a)$ is computed, for each value of $N$ and lattice spacing $a$, for several values of $\Rs$. The lower bound of the range is chosen to ensure $\Rs/a > 2.5$ (a reasonable choice to avoid UV contamination in the smoothened lattice topological charge density), while the upper bound to ensure $\Rs < \ell/2$ (with $\ell/2$ half the box size). With the adopted setup, these conditions correspond to $0.5 \lesssim \Rs\sqrt{\sigma} \lesssim 1.6$. After the subtraction of the singularity, the continuum limit is taken at fixed $\Rs^2 \sigma$, and assuming leading $\mathcal{O}(a^2\sigma)$ lattice artifacts. This procedure is shown in Fig.~\ref{fig:cont_limit}. Lattice artifacts are found to be very small and almost independent of $\Rs$ and $N$, and the continuum limit turns out to be smooth in all cases.

\begin{figure}[!t]
\centering                 
\includegraphics[scale=0.25]{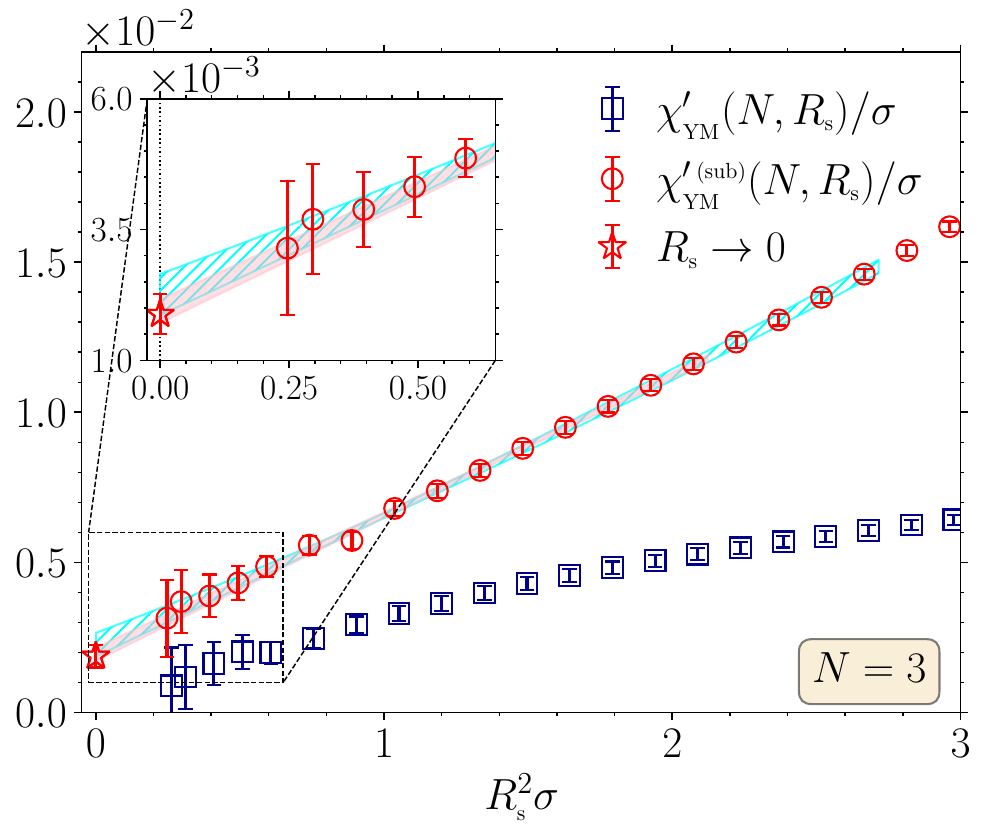}
\includegraphics[scale=0.25]{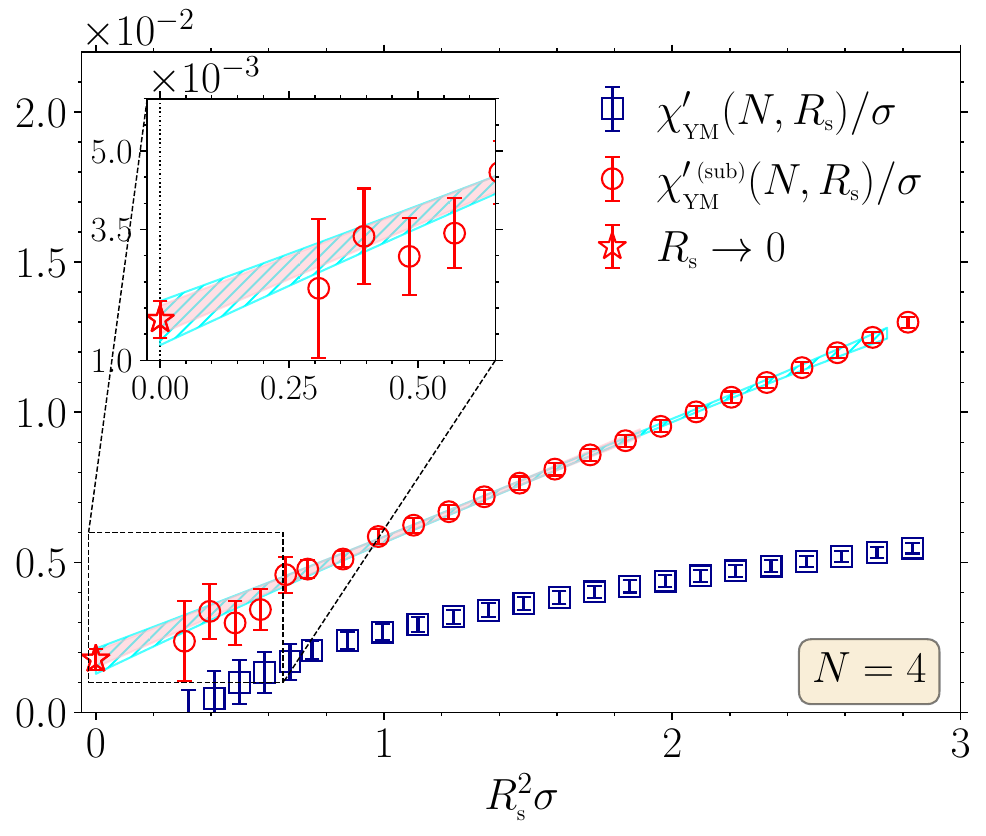}
\includegraphics[scale=0.25]{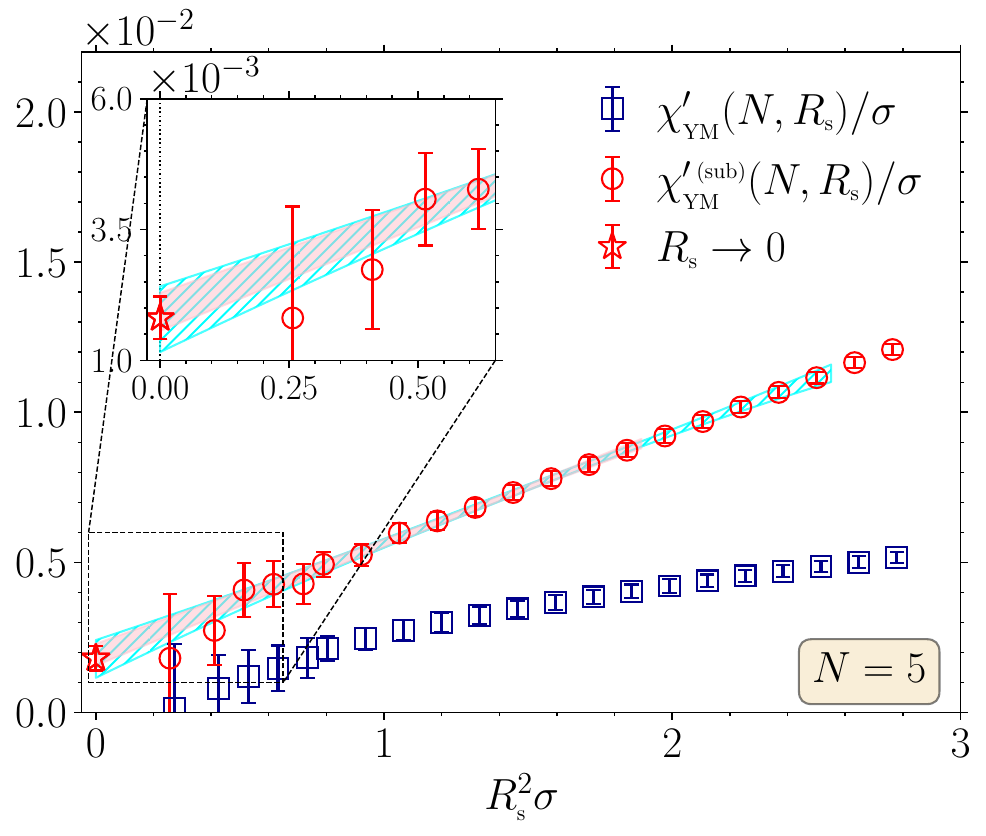}
\includegraphics[scale=0.25]{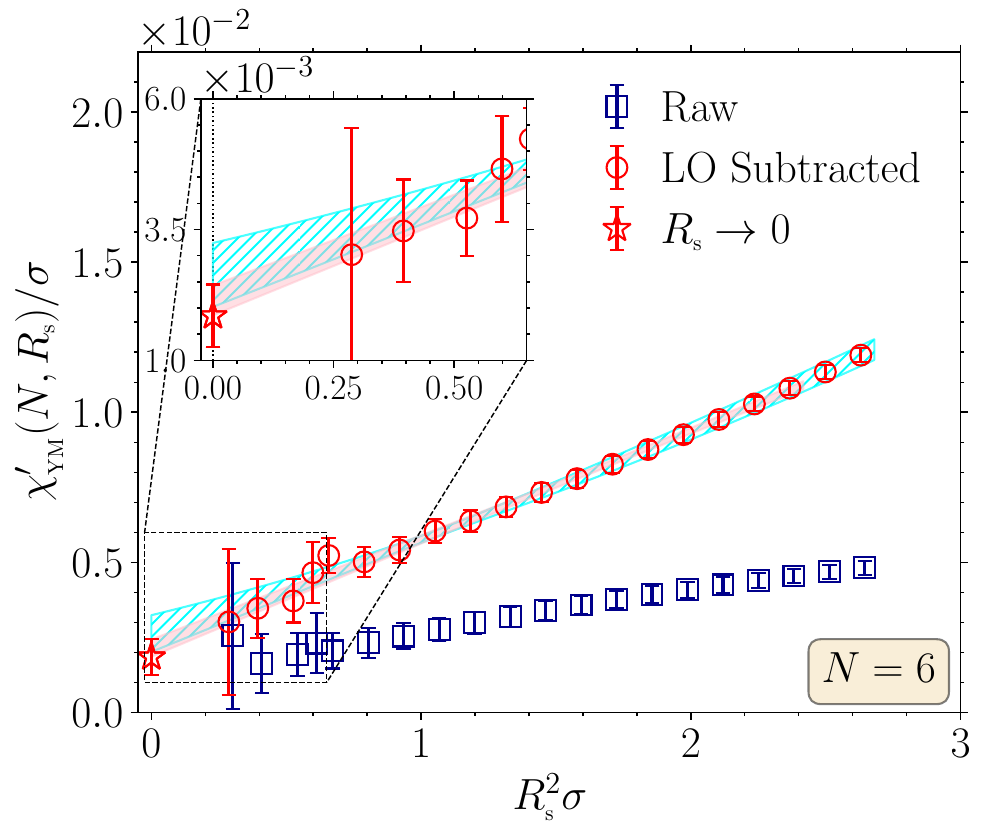}
\caption{Zero-$\Rs$ limits of $\chipsub(N,\Rs)/\sigma$ on selected bootstrap samples. Continuous/dashed bands represent examples of different $\Rs \to 0$ extrapolations (linear/quadratic) to check the fit stability. Star points in $\Rs^2\sigma=0$ are the final bootstrap-averaged results (including systematic errors).}
\label{fig:zeroRs_limit}
\end{figure}

As it was anticipated, $\chipsub$ exhibits, after the continuum limit, a non-trivial dependence on $\Rs$. Data are compatible with the theoretical expectation of a leading linear running of the subtracted susceptibility slope:
\bee\label{eq:zeroRs_fit_function}
\hspace*{-0.38\baselineskip}\chipsub(N,\Rs) = \chip_{_\YM}(N)\left[1 + M^2(N) \Rs^2 + \mathcal{O}(\Rs^4)\right].
\eee
The best fit of the continuum limits $\chipsub(N,\Rs)$ to Eq.~\eqref{eq:zeroRs_fit_function} allows the extraction of the scale-invariant susceptibility slope. This procedure is illustrated in Fig.~\ref{fig:zeroRs_limit}, and results are reported in Tab.~\ref{tab:final_res}.

The stability of the $\Rs \to 0$ extrapolation was explicitly checked for all values of $N$ by varying the upper bound of the linear fit range between $\Rs\sqrt{\sigma}\sim 0.9$ and $\sim 1.6$, and including/excluding a further $\mathcal{O}(\Rs^4)$ term in the fit function (found to be negligible in all cases). The resulting extrapolations are remarkably stable, as shown in the Supplemental Material, and systematic errors related to the spread of the extrapolations obtained with different fit Ans\"atze and fit ranges are negligible with respect to statistical uncertainties. This excellent fit stability suggests that neglected higher-order terms in the perturbative expansion of $S$ do not impact the $\Rs \to 0$ extrapolation from the employed range of $\Rs$ and with current errors.

\begin{figure}[!t]
\centering
\includegraphics[scale=0.5]{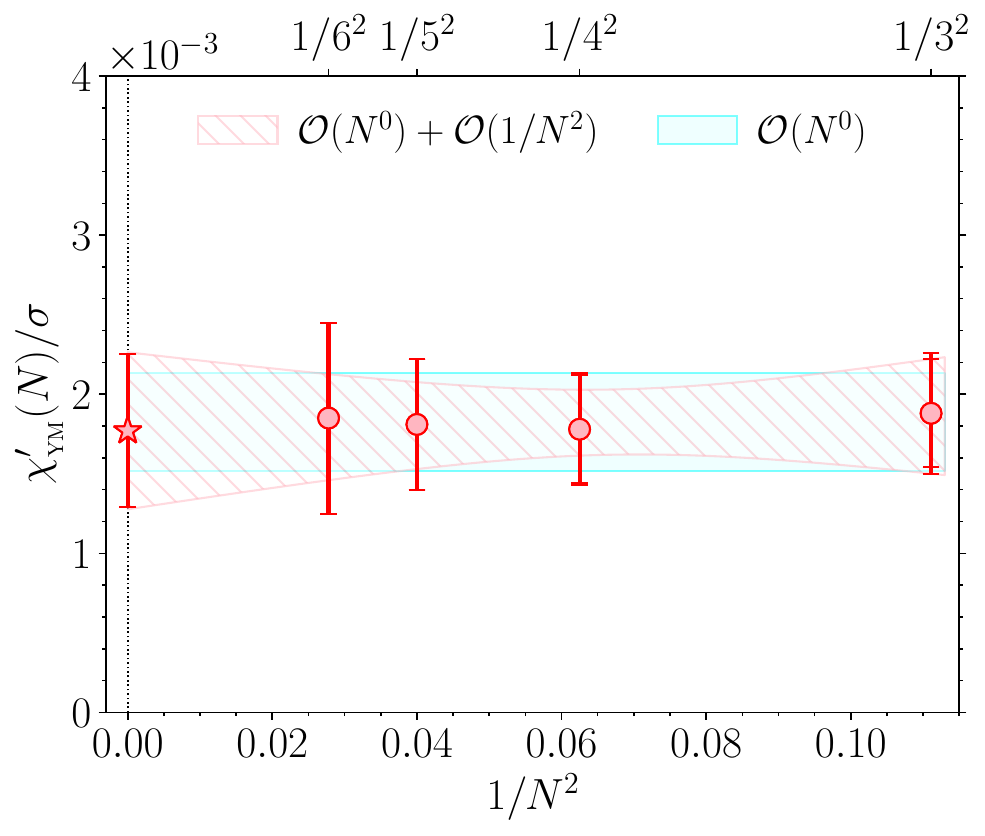}
\caption{Large-$N$ limit extrapolation of the scale invariant topological susceptibility slope $\chip_{_\YM}(N)/\sigma$.}
\label{fig:largeN_lim_chip}
\end{figure}

The scale-invariant susceptibility slope $\chip_{_\YM}(N)$ exhibits a smooth $N$-dependence, and the data support a finite large-$N$ limit, in agreement with the conjecture $\chip_{_\YM}(N)\sim\mathcal{O}(N^0)$ of~\cite{Giusti:2001xh}. In particular, data can be very well described by a behavior of the type:
\bee\label{eq:largeN_fit_function}
\chip_{_\YM}(N) = \chip_{_\YM}(\infty)\left[ 1 + C_2\frac{1}{N^2} + \mathcal{O}\left(\frac{1}{N^4}\right)\right].
\eee
The coefficient $C_2 = -0.5(3.9)$ of the $\mathcal{O}(1/N^2)$ term turns out to be consistent with zero within errors, and setting it to zero does not alter the good fit quality. This signals that sub-leading terms in the $1/N$ expansion cannot be resolved with current error bars. The difference in the extrapolation yielded by a linear and a constant fit is completely negligible, and the systematic error on the $N\to\infty$ value is dominated by the propagation of those on the $\Rs\to 0$ limit. Large-$N$ extrapolations of the scale-invariant susceptibility slope $\chip_{_\YM}(N)$ are shown in Fig.~\ref{fig:largeN_lim_chip}. A similar very mild $N$-dependence is found for the coefficient $M^2$ of the $\mathcal{O}(\Rs^2)$ term in $\chipsub(N,\Rs)$.

\begin{table}[!t]
\begin{tabular}{|c|c|c|}
\hline
\multicolumn{1}{|c|}{}&&\\[-1em]
\multicolumn{1}{|c|}{$N$} & $\frac{1}{\sigma}\chip_{_\YM}(N) \times 10^{3}$ & $\frac{1}{\sigma} M^2(N)$ \\
\multicolumn{1}{|c|}{}&&\\[-1em]
\multicolumn{1}{|c|}{}&&\\[-1em]
\hline
\multicolumn{1}{|c}{}&&\\[-1em]
3        & 1.88(34)$_{\rm stat}$(16)$_{\rm syst}$  & 2.57(45)$_{\rm stat}$(28)$_{\rm syst}$  \\
4        & 1.78(34)$_{\rm stat}$(06)$_{\rm syst}$  & 2.32(47)$_{\rm stat}$(03)$_{\rm syst}$  \\
5        & 1.81(41)$_{\rm stat}$(03)$_{\rm syst}$  & 2.18(54)$_{\rm stat}$(12)$_{\rm syst}$  \\
6        & 1.85(60)$_{\rm stat}$(06)$_{\rm syst}$  & 2.25(71)$_{\rm stat}$(04)$_{\rm syst}$  \\
$\infty$ & 1.77(48)$_{\rm stat}$(02)$_{\rm syst}$  & 2.03(62)$_{\rm stat}$(13)$_{\rm syst}$  \\
\hline
\end{tabular}
\caption{Final results for the scale invariant ($\Rs \to 0$) topological susceptibility slope and for the coefficient of the first sub-leading $\mathcal{O}(\Rs^2)$ correction, cf.~Eq.~\eqref{eq:zeroRs_fit_function}.}
\label{tab:final_res}
\end{table}

\section*{Discussion and conclusions}
\vs{-}

The obtained lattice results for $\chip_{_\YM}$ support the scenario where this quantity has the same large-$N$ scaling of the topological susceptibility. The final $N=\infty$ result in the scale-invariant limit reads (the total error is the quadrature sum of statistical and systematic ones):
\bee\label{eq:chip_YM_largeN}
\frac{1}{\sigma}\chip_{_\YM}(N=\infty) = 1.77(48)_{\rm stat+syst} \times 10^{-3},
\eee
remarkably close to the $N=3$ one,
\bee\label{eq:chip_YM_N3}
\frac{1}{\sigma}\chip_{_\YM}(N=3) = 1.88(38)_{\rm stat+syst} \times 10^{-3}.
\eee
Since Yang--Mills theories are not realized in Nature, the conversion of dimensionless quantities in physical units is a matter of conventions. To be conservative, several prescriptions have been chosen in order to assign a systematic error due to this scale-setting ambiguity (more details are in the Supplemental Material). In the end:
\beq\label{eq:chip_YM_NINF_PHYS}
\hspace{-\baselineskip}\chip_{_\YM}(N=\infty) &=& \left[20.3(2.8)_{\rm stat+syst}(1.0)_{\rm scale}\text{ MeV}\right]^2\\
&=& \left[20.3(2.9)\text{ MeV}\right]^2,
\eeq
\beq\label{eq:chip_YM_N3_PHYS}
\hspace{-\baselineskip}\chip_{_\YM}(N=3) &=& \left[20.9(2.1)_{\rm stat+syst}(1.0)_{\rm scale}\text{ MeV}\right]^2\\
&=& \left[20.9(2.4)\text{ MeV}\right]^2.
\eeq
Here, ``scale'' labels the systematic error associated to the large-$N$ scale-setting ambiguity, and is sub-leading compared to other sources of uncertainty. As a comparison, the QCD Sum Rule predicts $\chip_{_\YM}(N=3) \simeq \left[(7 \pm 3)\text{ MeV}\right]^2$. A useful way of cross-checking the plausibility of these lattice results for $\chip_{_\YM}$ comes from the internal consistency of the Witten--Veneziano mechanism. The Witten--Veneziano formula~\eqref{eq:witten_veneziano} assumes that
\beq
C_{_\YM}(p^2)= \chi_{_\YM} - \chip_{_\YM} p^2 + \mathcal{O}(p^4)
\eeq
is dominated by the $p^2=0$ term $\chi_{_\YM}$ up to $p^2 \sim m_{\etap}^2$. Thus, $\chip_{_\YM}$ should satisfy the bound~\cite{Narison:1991xxx,Narison:2006ws}:
\beq
\vert \chip_{_\YM} \vert \, m^2_{\etap} \ll \chi_{_\YM} \implies R\equiv\frac{\vert \chip_{_\YM} \vert m^2_{\etap}}{\chi_{_\YM}} \ll 1.
\eeq
In the strict large-$N$ limit, the bound is trivially satisfied since $R \sim \mathcal{O}(1/N)$, as the present results have shown that $\chip_{_\YM}\sim\mathcal{O}(N^0)$, while $\chi_{_\YM}\sim\mathcal{O}(N^0)$ and $m^2_{\etap}\sim\mathcal{O}(1/N)$.

Using Eq.~\eqref{eq:witten_veneziano} for $\Nf=3$, one finds $\chi_{_\YM}/m^2_{\etap}=F^2/6$, thus $R=6\chip_{_\YM}/F^2$. This expression can be evaluated using Eq.~\eqref{eq:chip_YM_largeN}, and the lattice large-$N$ result found in~\cite{Bonanno:2025hzr} via twisted Eguchi--Kawai reduction:
\bee\label{eq:Fpi_largeN}
F/\sqrt{N\sigma} = 0.1262(34) \qquad (N=\infty).
\eee
One finds that $R\ll 1$ already for $N=3$:
\bee
R = 0.67(18)/N \quad \implies \quad R\big\vert_{N\,=\,3} = 0.22(6).
\eee
These results further support the success of the Witten--Veneziano formula to describe the physical $\etap$ mass.

Combining the result that $\chip_{_\YM}\sim\mathcal{O}(N^0)$ and the lattice large-$N$ result in~\eqref{eq:Fpi_largeN}, one determines from Eq.~\eqref{eq:chipQCD_chipYM_formula}:
\bee\label{eq:chip_QCD_largeN_sigma}
\lim_{m\to0} \frac{1}{\sigma} \left(\frac{\chip_{_\QCD}}{N}\right)_{N\,=\,\infty} = -2.65(14)\times 10^{-3}.
\eee
Remarkably, it is found that $\chip_{_\QCD}$ in the chiral limit has the opposite sign compared to $\chip_{_\YM}$. This aspect is in qualitative agreement with the QCD Sum Rule, predicting $\chip_{_\QCD} < 0$ and $\chip_{_\YM} > 0$ too~\cite{Narison:2006ws,Narison:2021svo}. The reason for this change of sign could be due to the different balance between the finite contributions due to the $x=0$ positive contact term and the negative large-distance tail of $\braket{q(x)q(0)}$~\cite{Vicari:1999xx,Giusti:2001xh,Seiler:2001je}. On one hand, lowering the quark mass suppresses the QCD topological susceptibility with respect to the pure-gauge case (recall that $\chi$ is tightly related to the positive $x=0$ contact term of $\braket{q(x)q(0)}$, responsible for its strict positivity~\cite{Vicari:1999xx,Giusti:2001xh,Seiler:2001je}). On the other hand, light quark masses enhance the long-distance tail of the two-point correlator of $q(x)$, which in this limit decays with the $\etap$ mass, as opposed to a decay with the heavier pseudo-scalar glueball mass in the quenched limit.

Let us now discuss the implications of the result in Eq.~\eqref{eq:chip_QCD_largeN_sigma} for the Shore--Veneziano formula. From~\eqref{eq:chip_QCD_largeN_sigma}, one can obtain a leading $N=3$ determination as $\chip_{_\QCD}\big\vert_{N\,=\,3} = 3 \times (\chip_{_\QCD}/N)_{N\,=\,\infty}$:
\bee\label{eq:chip_QCD_largeN_chiral}
\lim_{m\to0}\chip_{_\QCD}= -\left[43.0(1.2)_{\sst{\text{$F$}}}(2.1)_{\sst{scale}}~\mathrm{MeV}\right]^2, \quad \left(\substack{\text{\tiny$N=3$ from}\\\text{\tiny$N=\infty$ result}}\right),
\eee
where the errors due to $F$ in Eq.~\eqref{eq:Fpi_largeN} and to the scale-setting ambiguity at large $N$ have been left separated. As a comparison, previous predictions from the QCD Sum Rule, the NJL model or $\chi$PT give all negative results ranging from $-(21\text{ MeV})^2$ down to $-(33\text{ MeV})^2$. Experimentally, the form factor $\gA$ is measured as a function of the exchanged photon virtuality $\mathcal{Q}^2$, and the running of $\chip_{_\QCD}(\Rs^2 \sim 1/\mathcal{Q}^2)$ is exactly necessary to reproduce the dependence of $\gA$ on $\mathcal{Q}^2$. Given that so far the scale-invariant limit has been discussed, it makes sense to consider the $\mathcal{Q}^2\to\infty$ extrapolation of the experimental data provided by COMPASS: $\gA(\mathcal{Q}^2\to\infty)=0.33(3)_{\sst{stat}}(5)_{\sst{syst}}=0.33(6)$~\cite{COMPASS:2006mhr}. The experimental measurement of $g_{\etap \sst{N}\sst{N}}$ is instead a much harder task. Currently, only the experimental bounds $0 \lesssim g_{\etap \sst{N}\sst{N}} \lesssim 2.5$ are available~\cite{Moskal:1998pc,Shore:2007yn}. Current theoretical estimates using various effective approaches are also affected by significant uncertainties, as different calculations yield results ranging from $\sim 0.4$ up to $\sim 8.6$, see the summary in~\cite{Singh:2018yvt} and references therein. Thus, a different route will be pursued here: using Eq.~\eqref{eq:shore_veneziano_formula}, the experimental value of $\gA$, and the non-perturbative determination of $\chip_{_\QCD}$, $g_{\etap\sst{N}\sst{N}}$ will be determined and compared with experimental bounds. The outcome of this procedure, adopting the latest PDG proton mass $m_{\sst{N}}=938.27$ MeV~\cite{ParticleDataGroup:2024cfk}, is:
\bee\label{eq:g_etapNN_SV}
g_{\etap \sst{N}\sst{N}} = 2.41(44)_{\sst{SV}}(12)_{\sst{scale}},
\eee
with SV the propagated error from Eq.~\eqref{eq:shore_veneziano_formula}. Since the Shore--Veneziano formula is obtained in large-$N$ QCD in the chiral limit, to compare Eq.~\eqref{eq:g_etapNN_SV} with experimental bounds I will assign it a further systematic error reflecting neglected finite-$N$/finite-$m$ corrections.

\begin{figure}[!t]
\centering
\includegraphics[scale=0.355]{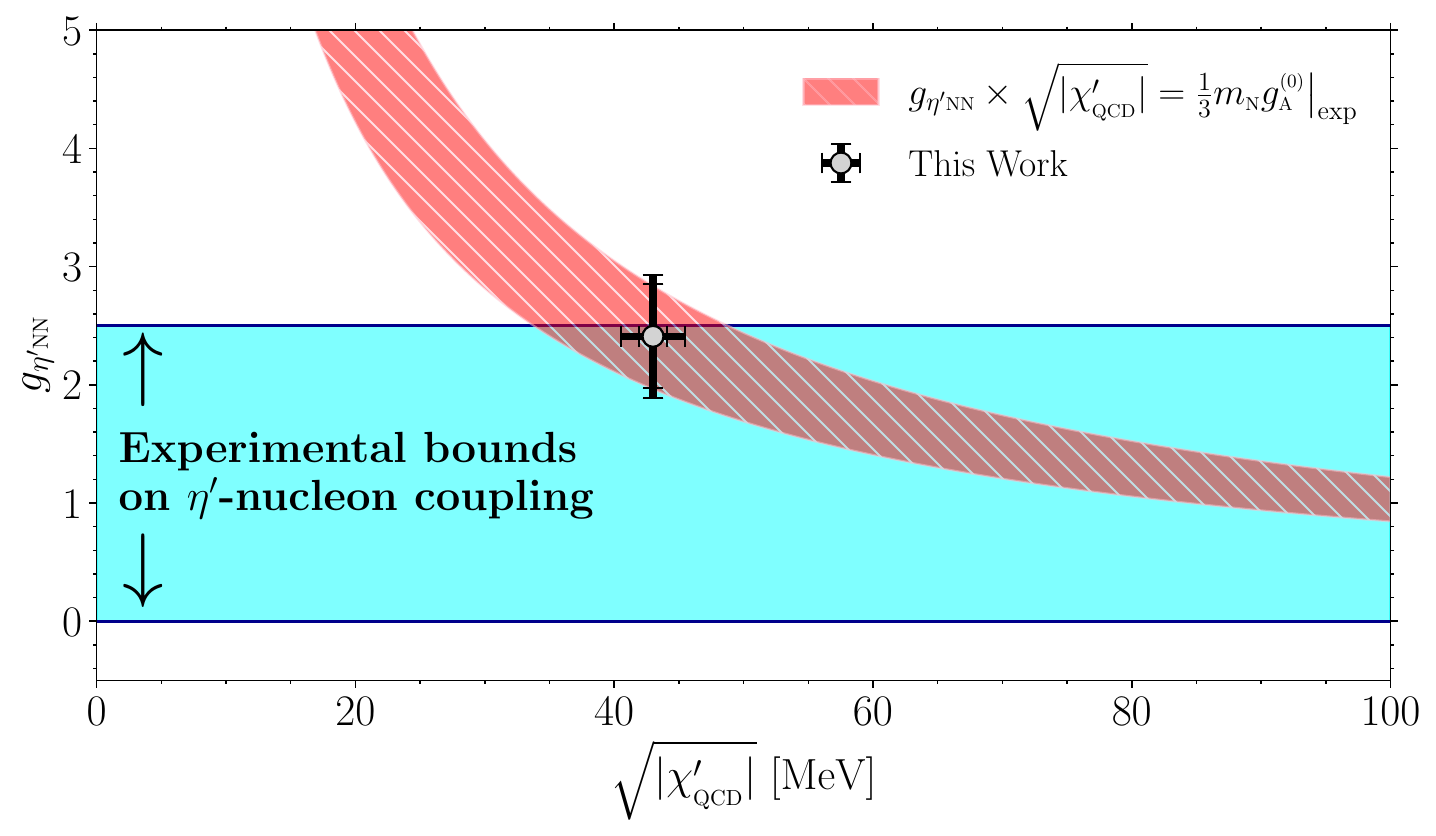}
\caption{Comparison of experimental bounds on $g_{\etap\sst{N}\sst{N}}$ with the determination obtained in this work via the Shore--Veneziano formula and the results for $\chip$ here presented.}
\label{fig:final}
\end{figure}

The starting point is the $\chi$PT prediction for $\chip_{_\QCD}$~\cite{Leutwyler:2000jg}:
\bee\label{eq:chip_QCD_chiPT}
\chip_{_\QCD} = - \frac{F^2}{2}\frac{\left(1+z_{\rm ud}^2+z_{\rm us}^2\right)}{\left(1+z_{\rm ud}+z_{\rm us}\right)^{2}}-\frac{H_0}{6}, \qquad \text{($\chi$PT)},
\eee
where $z_{\rm ud}=m_{\rm u}/m_{\rm d}$ and $z_{\rm us}=m_{\rm u}/m_{\rm s}$ are quark-mass ratios, and $H_0$ is an unknown low-energy constant representing the contact-term contribution to $\chip_{_\QCD}$. Taking the chiral limit of~\eqref{eq:chip_QCD_chiPT} with 3 degenerate flavors ($m_{\rm u}=m_{\rm d}=m_{\rm s}\equiv m\to 0$) and comparing it with Eq.~\eqref{eq:chipQCD_chipYM_formula}, one gets $-H_0/6 = \chip_{_\YM}$. Now, using Eq.~\eqref{eq:chip_QCD_chiPT}, Eq.~\eqref{eq:chip_YM_N3_PHYS}, and FLAG averages for $F$ and the quark masses~\cite{FlavourLatticeAveragingGroupFLAG:2021npn}, one obtains an estimate of $\chip_{_\QCD}$ in $N=3$ QCD with physical quark masses~\cite{suppmat}:
\bee
\chip_{_\QCD}= -\left[36.2(3.8)_{\sst{\text{QCD\,inputs}}}(0.6)_{\sst{scale}}~\mathrm{MeV}\right]^2, \quad \text{($\chi$PT)}.
\eee
This result can be used to repeat the calculation of $g_{\etap\sst{N}\sst{N}}$, and use the difference with respect to the result in Eq.~\eqref{eq:g_etapNN_SV} to quantify a systematic error due to neglected finite-$N$ and finite-$m$ corrections. The outcome is~\cite{suppmat}:
\bee\label{eq:g_etapNN_FINAL}
g_{\etap \sst{N}\sst{N}} = 2.41(44)_{\sst{SV}}(12)_{\sst{scale}}(25)_{\substack{\sst{finite\, \text{$N$}}\\\sst{finite\,\text{$m$}}}} = 2.41(52).
\eee
The comparison of the theoretical prediction for $g_{\etap\sst{N}\sst{N}}$ in Eq.~\eqref{eq:g_etapNN_FINAL} with experimental bounds is shown in Fig.~\ref{fig:final}. Remarkably, the obtained determinations for $g_{\etap\sst{N}\sst{N}}$ and $\sqrt{\vert\chip_{_\QCD}\vert}$ lie very close to the upper/lower bound (respectively) of the experimentally allowed region. These results support the Shore--Veneziano picture drawn in~\cite{Shore:2007yn}, connecting the suppression of $\gA$ to the smallness of $\sqrt{\vert\chip_{_\QCD}\vert}$. The prediction for $g_{\etap\sst{N}\sst{N}}$ is a non-trivial accomplishment of this work. It could be used to improve existing estimates of the model-independent QCD contribution to the axion-nucleon coupling~\cite{GrillidiCortona:2015jxo,DiLuzio:2021qct,Badziak:2023fsc,Candon:2025vpv}, and could provide useful theoretical guidance~\cite{Bass:2018xmz} to current experiments targeting its measurement~\cite{Skoupil:2025kmy,Akiyama:2025uyo}.

In the near future, it should be possible to reduce the error on~\eqref{eq:g_etapNN_FINAL}. On one hand, the future Electron-Ion Collider (EIC) will largely improve existing measurements of $\gA$~\cite{Abir:2023fpo}. On the other hand, precision lattice calculations of the quark~\cite{Barone:2026uyx} (see also~\cite{Green:2017keo,Alexandrou:2021wzv}) and gluon~\cite{CLQCD:2025dod} contributions to the proton spin have recently appeared. Thus, accurate first-principles results for $\gA$ are expected to become available soon.

A natural outlook of this study is to use the methods developed here to address the calculation of $\chip_{_\QCD}$ directly in real-world $N=3$ lattice QCD simulations with dynamical fermions, and study its quark mass dependence and its running with $\Rs$. This would allow to directly investigate the size of finite-$N$ corrections to the Shore--Veneziano formula, as well as to compare lattice data with $\chi$PT predictions for $\chip_{_\QCD}$. Another interesting future outlook would be to study $\chip$ at finite temperature. Given that the suppression of $\gA$ via $\chip_{_\QCD}$ is an effect due to the $\mathrm{U}(1)_{\sst{A}}$ anomaly, the possible high-$T$ effective $\mathrm{U}(1)_{\sst{A}}$ restoration should manifest in a drastic change of $\chip_{_\QCD}$ across the chiral crossover. Thus, the investigation of $\chip_{_\QCD}$ at finite temperature could provide a different viewpoint on the ongoing debate about effective $\mathrm{U}(1)_{\sst{A}}$ restoration in high-temperature QCD.

\vs{0.5}
{\footnotesize
\noindent \emph{\textbf{Acknowledgements.}} I gratefully acknowledge fruitful discussions with J.~L.~F.~Barb\'on, C.~Bonati, M.~D'Elia, M.~Garc\'ia P\'erez, N.~Husung, A.~Tarasov and R.~Venugopalan. I would also like to thank C.~Bonati, M.~D'Elia and M.~Garc\'ia P\'erez for reading a preliminary version of this manuscript, and for giving me useful suggestions to improve it. This work is partially supported by the Spanish Research Agency (Agencia Estatal de Investigación) through the grant IFT Centro de Excelencia Severo Ochoa CEX2020- 001007-S and, partially, by grant PID2021-127526NB-I00, both funded by MCIN/AEI/ 10.13039/ 501100011033. I also acknowledge support from the project H2020-MSCAITN-2018-813942 (EuroPLEx) and the EU Horizon 2020 research and innovation programme, STRONG-2020 project, under grant agreement No 824093. Numerical calculations have been performed on the \texttt{Finisterrae~III} cluster at CESGA (Centro de Supercomputaci\'on de Galicia).
}

\vs{0.5}
{\footnotesize
\noindent \emph{\textbf{Data Availability Statement.}} Data supporting the findings of this study are available in the Supplemental Material~\cite{suppmat}. Simulations have been performed via the open source code~\cite{PTBC}. Further data available upon reasonable request.
}

\renewcommand*{\bibfont}{\footnotesize}

\newpage
\clearpage
\onecolumngrid
\appendix

\setcounter{page}{1}
\section*{Supplemental Material}
\newcounter{count}
\Roman{count}
\setcounter{count}{1}

\setcounter{equation}{0}
\renewcommand{\theequation}{A.\arabic{equation}}

\setcounter{table}{0}
\renewcommand{\thetable}{A.\Roman{table}}

\setcounter{figure}{0}
\renewcommand\thefigure{A.\arabic{figure}}   

\section{A.\thecount~Matching Minkowski and Euclidean conventions for $\chi$ and $\chip$}

In the literature, different conventions for the definition of the topological susceptibility and the topological susceptibility slope have been employed. These are just the reflection of different conventions for the definition of the two-point function of the topological charge density and of its momentum expansion in Euclidean and in Minkowski space-times. Since this could be a source of confusion, the goal of this section is to match these different conventions in order to help the reader to compare this study with previous ones.

In this work, the Euclidean two-point function is defined as (all Euclidean quantities have the ``E'' subscript):
\beq
C_{\sst{E}}(p_{\sst{E}}^2) = \int \dd^4 x_{\sst{E}} \, \exp\left\{\ii p_{\sst{E},\mu} x_{\sst{E},\mu}\right\} \braket{q_{\sst{E}}(x_{\sst{E}})q_{\sst{E}}(0)}_{\sst{E}} = \chi_{_\sst{E}} - \chip_{_\sst{E}} p_{\sst{E}}^2 + \mathcal{O}(p^4_{\sst{E}}),
\eeq
where
\bee
\chi_{_\sst{E}} \equiv C_{\sst{E}}(0) = \int \dd^4 x_{\sst{E}} \, \braket{q_{\sst{E}}(x_{\sst{E}})q_{\sst{E}}(0)}_{\sst{E}}, \qquad \quad
\chip_{_\sst{E}} \equiv - \dfrac{\dd C_{\sst{E}}(p_{\sst{E}}^2)}{\dd p_{\sst{E}}^2} \bigg\vert_{p_{\sst{E}}^2 \, = \, 0} = \frac{1}{8}\int \dd^4 x_{\sst{E}} \, x_{\sst{E}}^2 \braket{q_{\sst{E}}(x_{\sst{E}})q_{\sst{E}}(0)}_{\sst{E}}.
\eee
With this convention, the topological susceptibility $\chi$ is \emph{strictly positive}, since $\chi_{_\sst{E}}=\braket{Q^2_{\sst{E}}}/V$, $Q_{\sst{E}}=\int \dd^4 x_{\sst{E}} \, q_{\sst{E}}(x_{\sst{E}})$, while the topological susceptibility slope $\chip_{_\sst{E}}$ has been determined in this work to be positive in Yang--Mills theories and negative in full QCD.

Conversely, the convention in Minkowski space-time is (here T is the time-ordered T-product):
\beq
C_{\sst{M}}(p^2) = \ii \int \dd^4x \exp\left\{\ii p_\mu x^\mu\right\} \braket{\mathrm{T}\left\{ q(x)q(0)\right\}} = \chi_{_\sst{M}} + \chip_{_\sst{M}} p^2 + \mathcal{O}(p^4),
\eeq
where
\beq
\chi_{_\sst{M}} \equiv C_{_\sst{M}}(0) = \ii \int \dd^4x \braket{\mathrm{T}\left\{ q(x)q(0)\right\}}, \qquad \quad
\chip_{_\sst{M}} \equiv \frac{\dd C_{_\sst{M}}(p^2)}{\dd p^2}\bigg\vert_{p^2\,=\,0} = -\frac{\ii}{8} \int \dd^4x \, x^2 \braket{\mathrm{T}\left\{ q(x)q(0)\right\}},
\eeq
see, e.g., Ref.~\cite{Tarasov:2025mvn}. In this convention, the quantities $\chi_{_\sst{M}}$ and $\chip_{_\sst{M}}$ are those that are called, respectively, topological susceptibility and topological susceptibility slope.

The goal is now to relate $\chi_{_\sst{M}}$ to $\chi_{_\sst{E}}$ and $\chip_{_\sst{M}}$ to $\chip_{_\sst{E}}$. Recalling the following rules for the Wick rotation,
\begin{align}
x^\mu = (x^0, \vec{x}), \quad p^\mu = (p_0, \vec{p}) & \quad \longrightarrow \quad x_{\sst{E},\mu} = (-\ii x_4, \vec{x}_{\sst{E}}), \quad p_{\sst{E},\mu} = (\ii p_4, -\vec{p}_{\sst{E}}),\\
\dd^4 x, \,\, x^2, \,\, p^2, \,\, p_\mu x^\mu & \quad \longrightarrow \quad -\ii \dd^4x_{\sst{E}}, \,\, -x_{\sst{E}}^2, \,\, -p_{\sst{E}}^2, \,\, x_{\sst{E},\mu} p_{\sst{E},\mu},\\
G_{0i}, \quad G_{ij} & \quad \longrightarrow \quad \ii G_{4i}, \quad G_{\sst{E},ij},\\
q(x) \propto \varepsilon_{\mu\nu\rho\sigma}G^{\mu\nu}G^{\rho\sigma} & \quad \longrightarrow \quad \ii q_{\sst{E}}(x_{\sst{E}})
\end{align}
it is easy to see that, after Wick rotation:
\beq
\chi_{_\sst{M}} \longrightarrow - \chi_{_\sst{E}}, \qquad \qquad \qquad \qquad \chip_{_\sst{M}} \longrightarrow -\chip_{_\sst{E}}.
\eeq
Therefore, with the Minkowski convention, all signs are the \emph{opposite} of the ones with the Euclidean convention. More precisely, $\chi_{_\sst{M}}$ is strictly \emph{negative}, while $\chip_{_\sst{M}}$ is \emph{negative} in the pure-gauge theory and \emph{positive} in full QCD.

\stepcounter{count}
\section{A.\thecount~Simulation details}

The Yang--Mills action was discretized on a hypercubic periodic lattice with volume $\ell^4 = (a L)^4$, with $a$ the lattice spacing and $L$ the number of lattice sites per direction, employing the standard Wilson discretization:
\begin{equation}
\mathcal{S}_{\sst{W}}[U] = -2Nb \sum_x \sum_{\mu \, > \, \nu}\Re\Tr\left[U_{\mu\nu}(x)\right],
\end{equation}
with $U_{\mu\nu}(x)=U_{\mu}(x)U_\nu(x+a\hat{\mu})U_\mu^\dagger(x+a\hat{\nu})U_\mu^\dagger(x)$ the plaquette rooted in site $x$ on the plane $(\mu,\nu)$, $b=1/\lambda=\beta/(2N^2)$ the bare inverse 't Hooft coupling, and $U_\mu(x)\in\SU(N)$ the gauge links. The overall scale was set using the string tension $\sigma$ via cubic spline interpolation of the results for $a^2\sigma$ of Refs.~\cite{Athenodorou:2020ani,Athenodorou:2021qvs}. For each $N$, a uniform range of lattice spacings was explored, $0.24\gtrsim a\sqrt{\sigma} \gtrsim 0.12$. This approximately corresponds to $0.10 \text{ fm} \gtrsim a \gtrsim 0.05 \text{ fm}$. The lattice size was taken to be approximately constant for all simulation points $\ell\sqrt{\sigma} \sim 3.4$ (about 1.5 fm). This choice was shown in~\cite{Bonanno:2023ple}, for $N=3$, to be sufficiently large to contain finite-size effects for all values of the smoothing radius considered here.

\begin{table}[!b]
\footnotesize
\begin{center}
\begin{tabular}{|c|c|c|c|c|c|c|c|c|c|c|c|}
\hline
&&&&&&&&&&&\\[-1em]
$N$ & $\beta$ & $b=\dfrac{1}{\lambda}$ & $L$ & $a\sqrt{\sigma}$ & \makecell{$\ell\sqrt{\sigma}$\\ $= a L \sqrt{\sigma}$} & \makecell{$n_\cool$\\max} & \makecell{$\Rs \sqrt{\sigma}$\\max} & $n_{\scriptscriptstyle{\rm meas}}$ & $N_{\rep}$ & $L_{\defect}$ & \makecell{$\ell_{\defect}\sqrt{\sigma}$ \\ $= a L_{\defect} \sqrt{\sigma}$}\\
\hline
\multicolumn{12}{c}{}\\[-1em]
\cline{1-9}
\multirow{5}{*}{3} & 5.95 & 0.3306 & 16 & 0.23567(69) & 3.77 & 19 & 1.678 & 1.94M & \multicolumn{3}{c}{} \\
& 6.00 & 0.3333 & 16 & 0.21609(76) & 3.46 & 22 & 1.654 & 2.03M & \multicolumn{3}{c}{} \\
& 6.07 & 0.3372 & 18 & 0.19238(51) & 3.46 & 28 & 1.662 & 1.94M & \multicolumn{3}{c}{} \\
& 6.20 & 0.3444 & 22 & 0.15788(31) & 3.47 & 42 & 1.668 & 1.23M & \multicolumn{3}{c}{} \\
& 6.40 & 0.3556 & 30 & 0.11879(26) & 3.57 & 75 & 1.682 & 0.71M & \multicolumn{3}{c}{} \\
\cline{1-9}
\multicolumn{12}{c}{}\\[-1em]
\cline{1-9}
\multirow{4}{*}{4} & 11.02 & 0.3444 & 16 & 0.21434(28) & 3.43 & 21 & 1.679 & 2.08M & \multicolumn{3}{c}{} \\
\cline{10-12}
& 11.20 & 0.3500 & 20 & 0.18149(49) & 3.63 & 32 & 1.677 & 198k & 16 & 3 & 0.54 \\
& 11.40 & 0.3563 & 24 & 0.15305(34) & 3.67 & 45 & 1.677 & 206k & 16 & 3 & 0.46 \\
& 11.60 & 0.3625 & 26 & 0.13065(21) & 3.40 & 64 & 1.706 & 262k & 24 & 4 & 0.52 \\
\hline
\multicolumn{12}{c}{}\\[-1em]
\hline
\multirow{4}{*}{5} & 17.43  & 0.3486 & 16 & 0.22217(37) & 3.55 & 21 & 1.663 & 777k & 12 & 2 & 0.44 \\
& 17.63  & 0.3526 & 18 & 0.19636(35) & 3.53 & 28 & 1.697 & 464k & 20 & 3 & 0.59 \\
& 18.04  & 0.3608 & 22 & 0.15622(38) & 3.44 & 42 & 1.653 & 380k & 20 & 3 & 0.47 \\
& 18.375 & 0.3675 & 26 & 0.13106(30) & 3.41 & 60 & 1.658 & 118k & 32 & 4 & 0.52 \\
\hline
\multicolumn{12}{c}{}\\[-1em]
\hline
\multirow{4}{*}{6} & 25.32 & 0.3517 & 16 & 0.22208(35) & 3.55 & 20 & 1.622 & 611k & 14 & 2 & 0.44 \\
& 25.70 & 0.3569 & 18 & 0.18956(33) & 3.41 & 28 & 1.638 & 325k & 24 & 3 & 0.57 \\
& 26.22 & 0.3642 & 22 & 0.15480(36) & 3.41 & 42 & 1.638 & 200k & 24 & 3 & 0.46 \\
& 26.65 & 0.3701 & 26 & 0.13173(29) & 3.42 & 57 & 1.624 & 115k & 36 & 4 & 0.53 \\
\hline
\end{tabular}
\end{center}
\caption{Summary of simulation parameters. For each point both the standard inverse coupling $\beta=2N/g^2$ and the inverse 't Hooft coupling $b=1/\lambda=1/(Ng^2)=\beta/(2N^2)$ (with 4 significant digits) are reported. The quantity $n_{\scriptscriptstyle{\rm meas}}$ is the number of samples of $a^2 \chip_{_\YM}$, always measured every 10 updating steps (except for $N=3$, $\beta=6.40$, where measurements are separated by 100 updates).}
\label{tab:simulation_summary}
\end{table}

The lattice topological charge density is discretized via the standard clover formulation:
\beq
q_{_\L}(x) = \frac{1}{32\pi^2} \sum_{\mu\nu\rho\sigma} \varepsilon_{\mu\nu\rho\sigma} \Tr\left[\mathcal{C}_{\mu\nu}(x)\mathcal{C}_{\rho\sigma}(x)\right],
\eeq
i.e., the simplest parity-odd lattice formulation of $q(x)$. Here $\mathcal{C}_{\mu\nu}(x)$ is the imaginary part of the average of the 4 plaquettes containing the site $x$ and lying in the $(\mu,\nu)$ plane. Wilson cooling used to smooth the gauge configurations entering the calculation of the lattice clover topological charge density is implemented iteratively: at each cooling step, each link is aligned to its local force, computed according to the Wilson plaquette action. In Ref.~\cite{Bonanno:2025eeb}, it was shown that, for all choices of the smoothing radius employed in this study, the continuum limit of $\chi_{_\YM}$ was independent of $\Rs$, as expected on general theoretical grounds, and in perfect agreement with previous studies employing coarser lattice spacings to take the continuum limit.

Concerning the Monte Carlo updating algorithm, for the simulations for which topological freezing is not an issue (all $N=3$ simulations and the coarsest lattice spacing of $N=4$), the customary 4:1 mixture of Cabibbo--Marinari heat-bath and over-relaxation was employed. The Parallel Tempering on Boundary Conditions (PTBC) algorithm was instead employed for all $N>3$ simulations (except for the coarsest $N=4$ point). One considers $N_\rep$ replicas of the lattice, differing for the boundary conditions imposed on the temporal links stemming from a cubic defect with size $\ell_\defect = a L_\defect \sim 0.5/\sqrt{\sigma}$, i.e., $\ell_\defect \sim 0.2$ fm, placed on the temporal boundary. Boundary conditions are altered changing $b \to b \times c(r)$ along the defect, with $0 \le c(r) \le 1$ (with the extremes corresponding to $1=$ periodic, $0=$ open). Neighboring replicas are allowed to swap gauge configurations at equilibrium via a standard Metropolis test. The boundary condition tempering parameters $c(r)$ are tuned through short runs to ensure an almost fixed swap rate $p\simeq 20\%$. Translations of the periodic replica are alternated with swaps to create new topological excitations around the periodic lattice. All measurements are performed on the periodic replica free of systematic effects from the open boundaries. With this setup, the number of replicas $N_r$ is found to scale as $N_\rep \sim 0.8 N L_\defect^{-3/2}$, while the autocorrelation time of $Q$ in the periodic replica is found to scale as $\tau_Q\sim 0.16 \sqrt{N} /(a^2 \sigma)$.

The gauge configurations employed in this study are inherited from Ref.~\cite{Bonanno:2025eeb}, where the large-$N$ limit of the Yang--Mills topological susceptibility was studied. Their simulation parameters are summarized in Tab.~\ref{tab:simulation_summary}. Additional simulations performed to check finite-size effects and to compare cooling with gradient flow will be described in the following sections.

\stepcounter{count}
\section{A.\thecount~The topological susceptibility slope, topological freezing, and finite-volume effects}

The topological susceptibility slope $\chip$, although depending on local fluctuations of the topological charge density $q(x)$, turns out to be strongly correlated with the global topological background of the gauge configuration $Q=\int \dd^4 x \, q(x)$. This implies that a sampling problem in the global topological charge due to topological freezing would introduce a large bias in $\chip$. This is expected to manifest as large power-like finite-volume effects~\cite{Brower:2003yx}. Thus, without a definite strategy to deal with topological freezing, prohibitively large volumes would be needed to remove the bias.

A straightforward way to visualize how strongly $\chip$ is correlated with the global topological background, is to compute the following projected quantity:
\bee\label{eq:chip_projQ}
a^2\chip_{_\YM}(Q) = \frac{1}{8} \sum_{x} d^2(x,0) \frac{\braket{q(x)q(0) \delta_Q}_{_\YM}}{\braket{\delta_Q}_{_\YM}},
\eee
as opposed to the actual observable $a^2\chip_{_\YM} = \frac{1}{8} \sum_{x} d^2(x,0)\braket{q(x)q(0)}_{_\YM}$. Here, $d(x,0)$ represents the shortest Euclidean distance (with periodic boundary conditions) between sites $x$ and $0$
\bee
d^2(x,y) = \sum_\mu d^2_\mu(x,y), \qquad d^2_\mu(x,y) =
\begin{cases}
(x_\mu-y_\mu)^2,   & \text{if } \vert x_\mu - y_\mu \vert \le \ell/2, \\
[\ell - \vert x_\mu-y_\mu\vert]^2,    & \text{if } \vert x_\mu - y_\mu \vert > \ell/2.
\end{cases}
\eee
The projector $\delta_Q$ is equal to 1 if the rounded lattice topological charge is equal to $Q$, and to 0 otherwise.

\begin{figure}[!htb]
\centering
\includegraphics[scale=0.34]{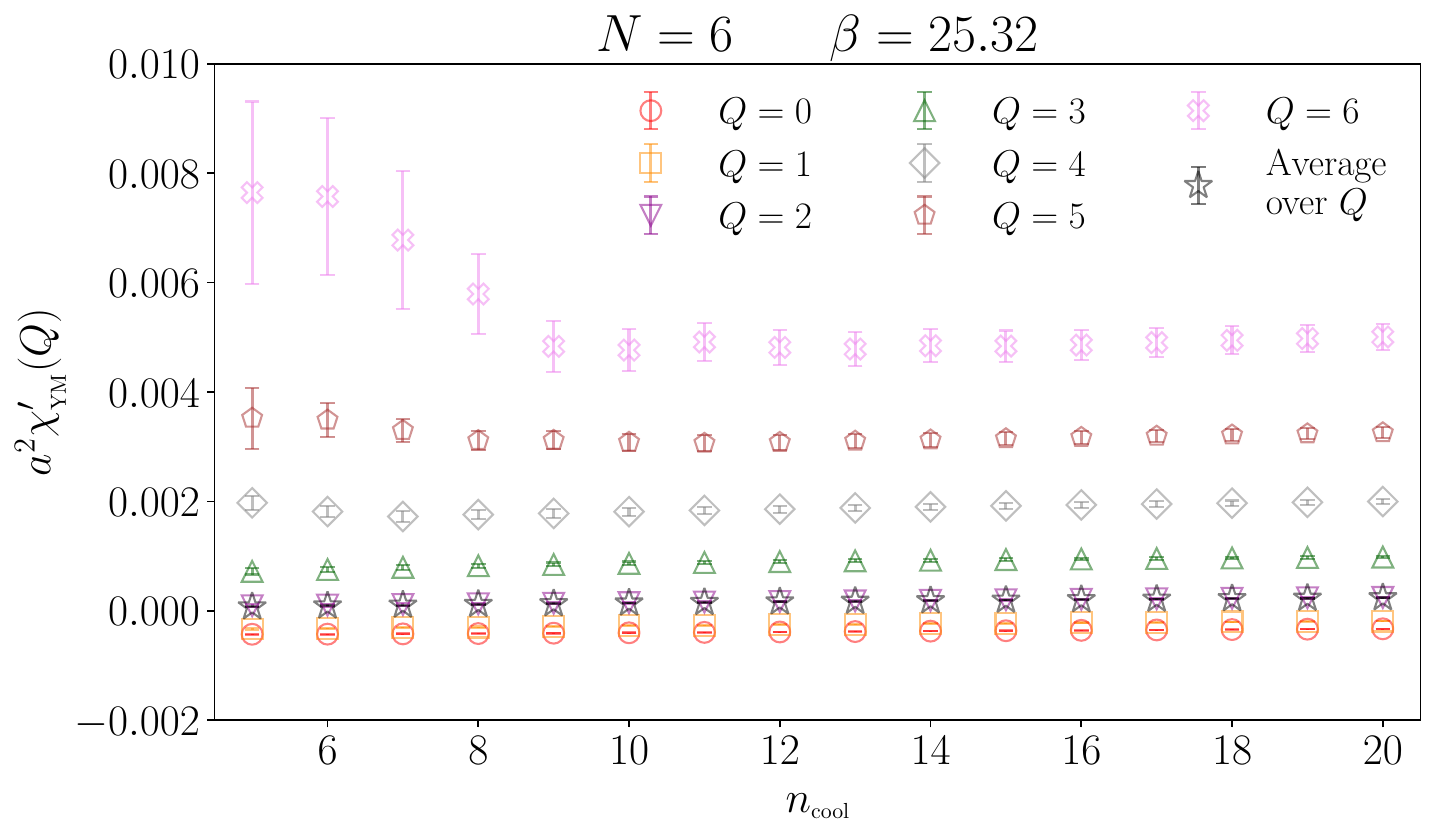}
\includegraphics[scale=0.34]{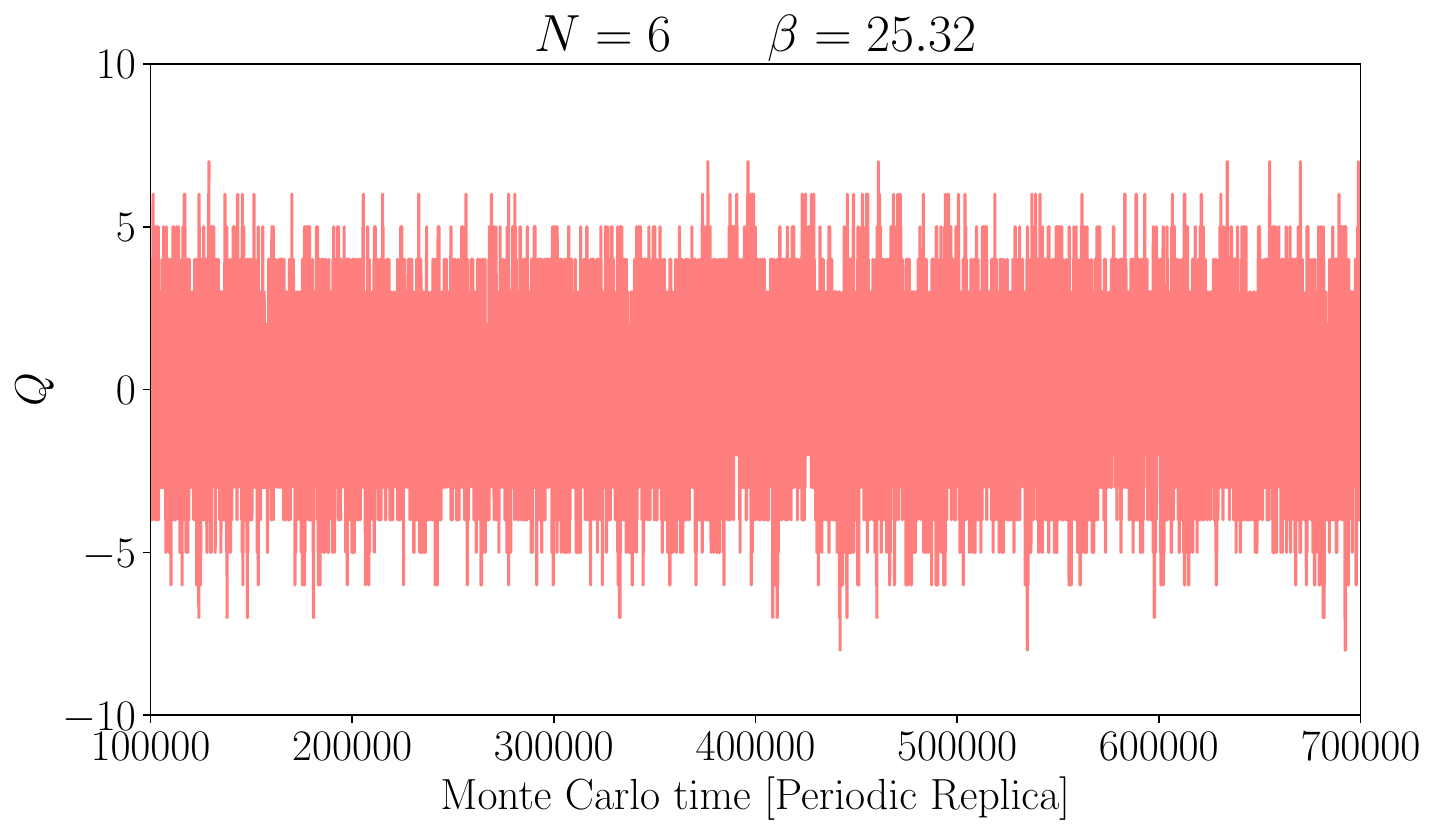}
\includegraphics[scale=0.34]{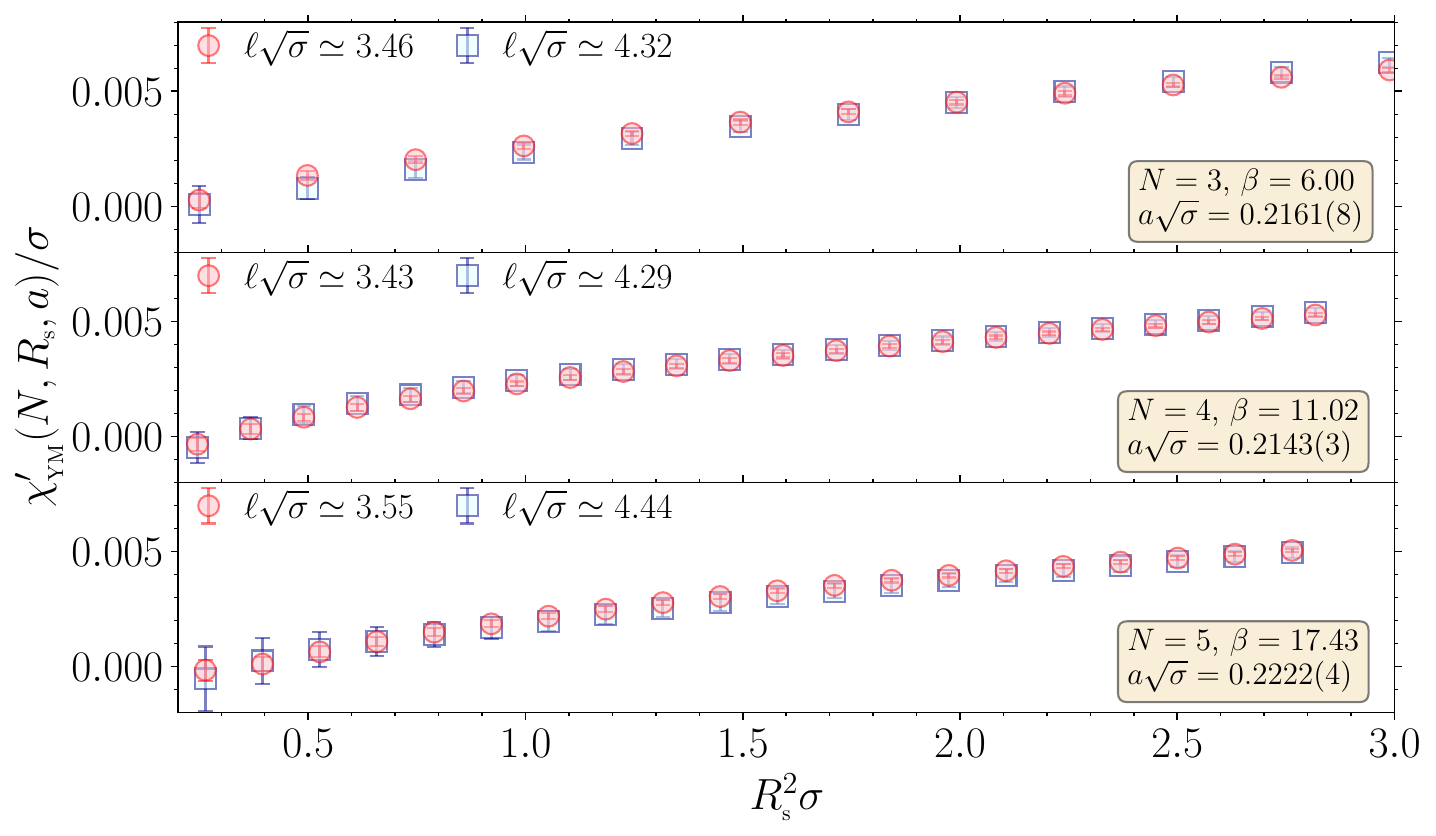}
\caption{Top left panel: example of the dependence of $\chip_{_\YM}$ on the topological sector, computed according to Eq.~\eqref{eq:chip_projQ}. Top right panel: PTBC Monte Carlo evolution of the rounded lattice topological charge ($n_{\cool}=20$) (10\% of the total statistics). Bottom plot: study of finite-size effects for $N=3,4,5$ with additional simulations performed for a larger volume.}
\label{fig:chip_projQ}
\end{figure}

After cooling, $Q_{_\L}= \sum_x q_{_\L}(x)$ is always very close to an integer number, thus an integer lattice topological charge can be obtained rounding $Q_{_\L}$ to the nearest integer following the procedure of Refs.~\cite{Bonati:2015sqt,Durr:2025qtq}. An example of $a^2\chip_{_\YM}(Q)$ as a function of $n_\cool$ and for $Q=0,1,\dots,6$ is shown in Fig.~\ref{fig:chip_projQ} (top left panel). The corresponding Monte Carlo evolution of $Q$ is shown in the top right panel of Fig.~\ref{fig:chip_projQ}. Since $\chip_{_\YM}$ exhibits a strong dependence on $Q$, topological freezing is expected to introduce large power-like finite-volume effects in this quantity. Thanks to the PTBC algorithm, these are avoided. The absence of significant finite-volume effects was explicitly checked performing additional simulations for $N=3,4,5$ and a lattice spacing $a\sqrt{\sigma}\approx 0.22$ on a larger volume. The result is illustrated in the bottom panel of Fig.~\ref{fig:chip_projQ}. Determinations of $\chip_{_\YM}$ obtained for $\ell\sqrt{\sigma} \approx 3.4$ (the lattice size employed in the main study) and $4.3$ are perfectly consistent within errors in the explored range of smoothing radii $\Rs^2\sigma < 3$.

\stepcounter{count}
\section{A.\thecount~Singular-term subtraction}

In Ref.~\cite{Giusti:2001xh}, it was shown that the lattice topological charge density correlator (defined in terms of an overlap-based discretization of $q$) possesses a contact-term in the origin that contributes up to $\mathcal{O}(p^4)$. In particular, at $\mathcal{O}(p^2)$, this would result in a $\mathcal{O}(p^2/a^2)$ divergent term in the continuum limit $a\to 0$. When defining the correlation function with a flowed gluonic definition of $q$, such divergent term is expected to be regularized, and the flowed correlator is expected to have a finite continuum limit for every $t>0$. However, such contact term may show up again (by power counting) as a $p^2/t$ divergence in the limit $t\to 0$. Based on this line of reasoning, one can expect a $1/t$ divergence in $\chip$ when taking the small-flow-time limit $t\to 0$. The goal of this section is to analyze this issue at leading order in perturbation theory.

The starting point is the computation of the topological charge density correlator at leading order in perturbation theory defining the topological charge density from gauge fields at positive flow time $t>0$. This calculation is expected to be reliable in the small flow time limit $t\to 0$. The procedure followed here closely resembles the one carried on in Ref.~\cite{Luscher:2010iy}, and the same conventions and notation used in that study will be also used here.

Truncating the perturbative expansion of the flowed field $B_\mu = g_0 B_\mu^{\one} + \mathcal{O}(g_0^2)$ at leading order ($g_0$ is the bare coupling), and choosing the gauge-fixing condition $\lambda_{\rm gf}=1$, one obtains the following relation~\cite{Luscher:2010iy}:
\bee\label{eq:flowed_field_LO}
B^{\one}_\mu(p,t) \simeq \ee^{- tp^2}A_\mu(p), \qquad \text{where} \qquad O(p) = \int \dd^4 x \, \ee^{-\ii p \cdot x} \, O(x).
\eee
At this order of the perturbative expansion, the flowed topological charge density reads:
\bee\label{eq:qt_LO}
q_t(x) =  \frac{g_0^2}{16\pi^2} \varepsilon_{\mu\nu\rho\sigma}\partial_\mu B^{\one a}_{\nu}(x,t) \partial_\rho B^{\one a}_{\sigma}(x,t) + \mathcal{O}(g_0^3) \equiv c_q \varepsilon_{\mu\nu\rho\sigma}\partial_\mu B^{\one a}_{\nu}(x,t) \partial_\rho B^{\one a}_{\sigma}(x,t) + \mathcal{O}(g_0^3),
\eee
with $c_q=[g_0/(4\pi)]^2$, and where $2\Tr[T_a T_b]=\delta_{ab}$ has been used. Introducing the flowed correlator in coordinate space $C_t(x)$, and the flowed correlator in momentum space $C_t(p^2)$:
\beq
C_t(x) = \braket{q_t(x)q_t(0)}, \qquad \qquad C_t(p^2) &=& \int \dd^4 x\, \ee^{\ii p \cdot x} C_t(x),
\\
\label{eq:Ctp2_def}
\braket{q_t(p)q_t(p^\prime)} &=& (2\pi)^4 \delta^{\sst{(4)}}(p+p^\prime)\,C_t(p^2),
\eeq
one can use Eq.~\eqref{eq:qt_LO} and the expression of $B_\mu^{\one}$ in terms of $A_\mu$ in Eq.~\eqref{eq:flowed_field_LO} to evaluate the left-hand side of Eq.~\eqref{eq:Ctp2_def} at leading order. Once $C^{\sst{(\LO)}}_t(p^2)$ is obtained, one can expand it at $\mathcal{O}(p^2)$ and finally obtain $-\chip_{_\LO}(t)$.

After performing the relevant Wick contractions, and using the Feynman-gauge gluon propagator $\langle A_\mu^a(p)A_\nu^b(p^\prime)\rangle_0 = (2\pi)^4\delta^{(4)}(p+p')\,\delta^{ab}\frac{\delta_{\mu\nu}}{p^2}$, one finally arrives to the following expression:
\beq
C_t^{\sst{(LO)}}(p^2) &=& 4 c_q^2 d_{\sst{A}}\int_k \ee^{-2t(k^2+q^2)} \frac{k^2 q^2-(k\cdot q)^2}{k^2 q^2} + \mathcal{O}(g_0^6)\\
&=&\frac{g_0^4(N^2-1)}{64 \pi^4}\int_k \ee^{-2t(k^2+q^2)} \frac{k^2 q^2-(k\cdot q)^2}{k^2 q^2} + \mathcal{O}(g_0^6), \qquad \qquad q = p-k, \qquad \int_p \equiv \int \frac{\dd^4 p}{(2\pi)^4},
\eeq
where $d_{\sst{A}} = N^2 - 1$ has been used. Interestingly, $C_t^{\sst{(LO)}}(p^2=0)=\chi_{_\LO}=0$, in agreement with the well-known result that the topological susceptibility is zero at all orders in perturbation theory. At $\mathcal{O}(p^2)$, one has
\bee
C_t^{\sst{(LO)}}(p^2) = \frac{g_0^4(N^2-1)}{64 \pi^4}\int_k \ee^{-4tk^2} \frac{k^2 p^2-(k\cdot p)^2}{k^4} + \mathcal{O}(p^4,g_0^6),
\eee
that, by rotational invariance, reduces to:
\bee
C_t^{\sst{(LO)}}(p^2) = \frac{g_0^4(N^2-1)}{64 \pi^4}  \frac{3}{4} p^2 \int_k \ee^{-4tk^2} \frac{1}{k^2} + \mathcal{O}(p^4,g_0^6),
\eee
with the remaining integral being equal to $1/(64\pi^2 t)$. Thus, recalling the definition $\chip=-C^\prime(p^2)\vert_{p^2 \, = \, 0}$, one finally arrives to:
\bee
\chip_{_\LO}(t) = - \frac{3(N^2-1)}{2^{14} \pi^6} \frac{g^4(\mu_t)}{t}[1+\mathcal{O}(g^2)],
\eee
where the bare coupling $g_0$ has been substituted with the running coupling $g(\mu)$ at the scale $\mu=\mu_t=1/\sqrt{8t}$. Note that $\chip_{_\LO}(t)\sim\mathcal{O}(N^0)$ since, at large $N$, $(N^2-1)g^4 \sim \lambda^2\sim\mathcal{O}(N^0)$, with $\lambda$ the 't Hooft coupling. Remarkably, in perturbation theory $\chip_{_\LO}(t)$ exactly takes the form of a $1/t$ singularity, as discussed in the beginning of this section. This singular term is exactly the one that is subtracted prior to define the subtracted susceptibility slope employed in the analysis presented in the main text, cf.~Eq.~\eqref{eq:chip_subtraction_prescription}. Apart from this singularity, for $x\ne 0$ one has $C_t(x) = C_{t\,=\,0}(x) + \mathcal{O}(t)$~\cite{Ce:2015qha}, thus, the regular part of the correlator only introduces $t\to 0$ vanishing corrections.

For the purpose of a numerical evaluation of $\chip_{_\LO}(t)$, a one-loop or two-loop approximation of the running coupling would be too crude. Instead, I have opted for a definition of the coupling based on the action density. Recalling that~\cite{Luscher:2010iy},
\bee
\mathcal{E}(t) \equiv t^2 \braket{E(t)} = \frac{3(N^2-1)}{2^7 \pi^2} g^2(\mu_t)[1 + \mathcal{O}(g^2)], \qquad E = \frac{1}{4}G_{\mu\nu}^aG_{\mu\nu}^a,
\eee
one can trade $g^2(\mu_t)$ for $\mathcal{E}(t)$. Doing so leads to:
\bee
\chip_{_\LO}(t) = -\frac{1}{3\pi^2} \frac{\mathcal{E}^2(t)}{(N^2-1)t} [1 + \mathcal{O}(\mathcal{E})]
\eee

\begin{figure}[!t]
\centering
\includegraphics[scale=0.25]{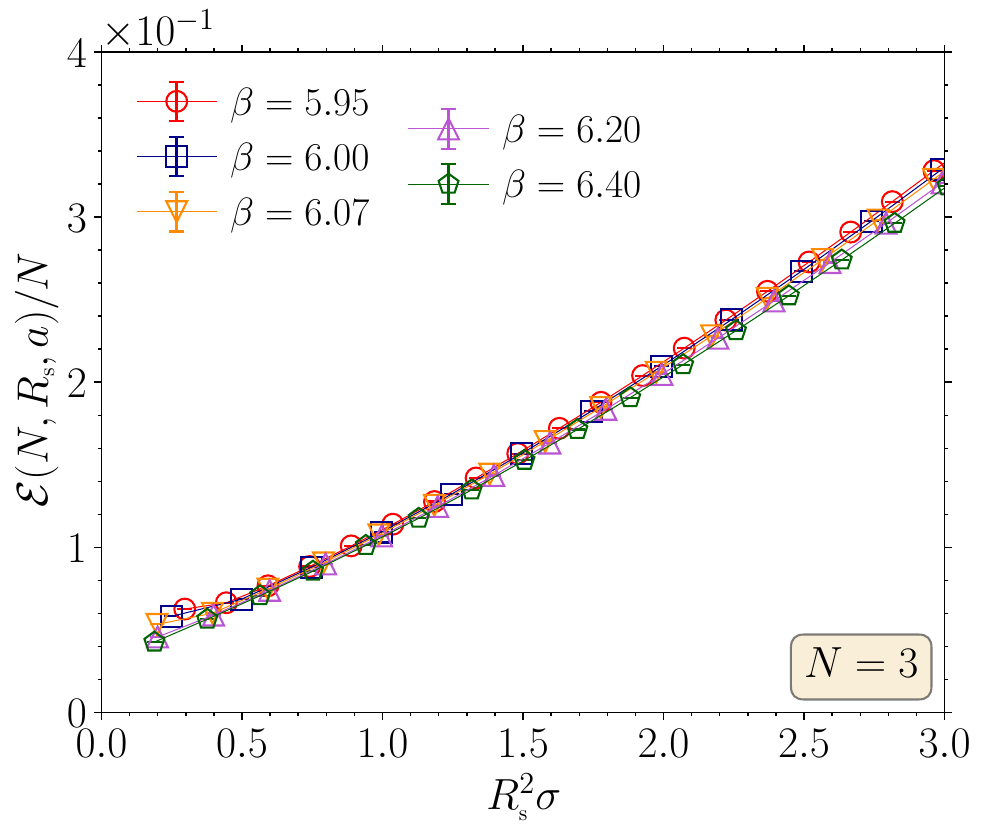}
\includegraphics[scale=0.25]{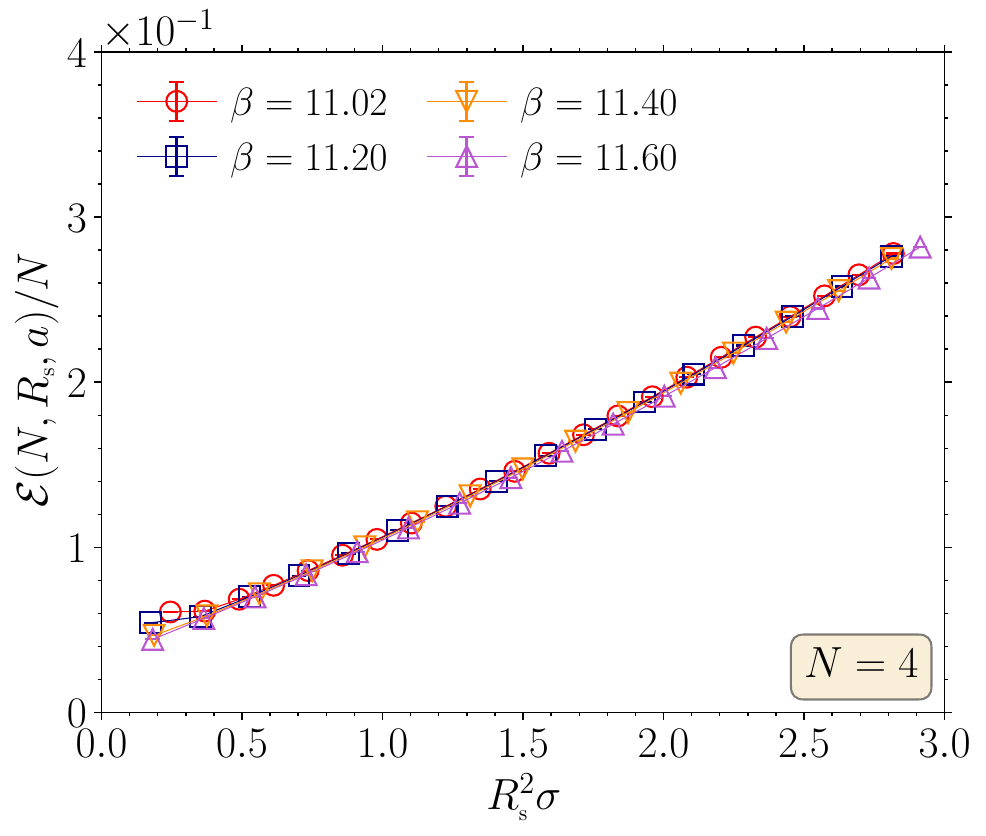}
\includegraphics[scale=0.25]{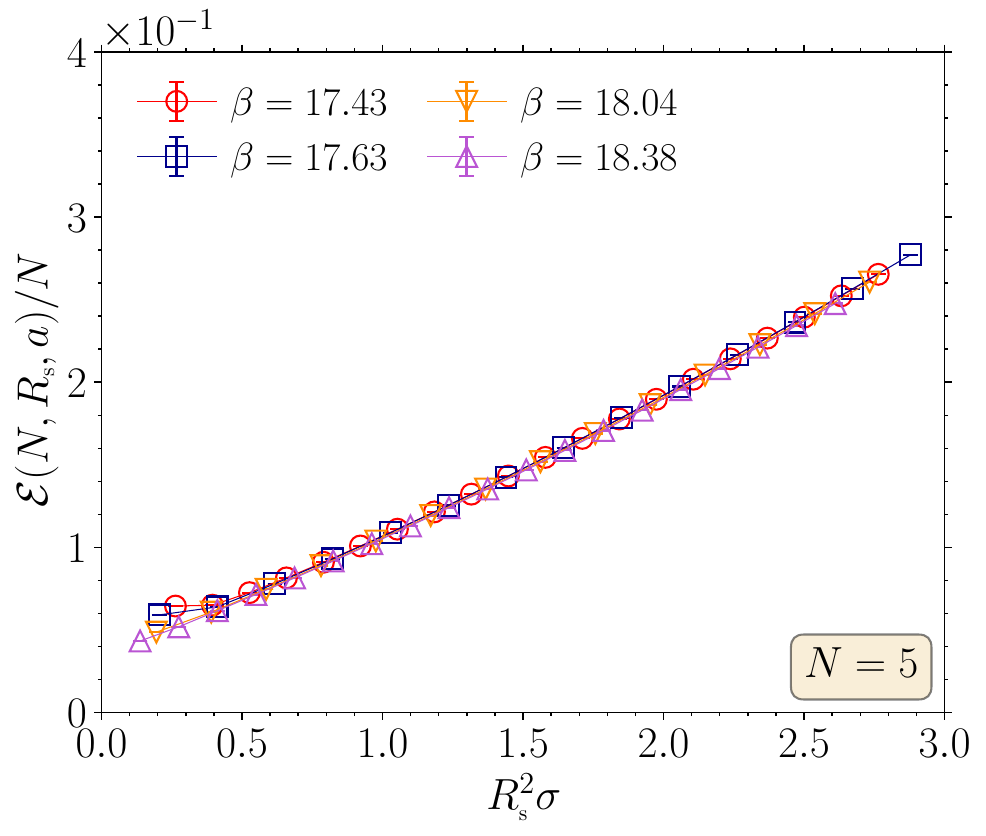}
\includegraphics[scale=0.25]{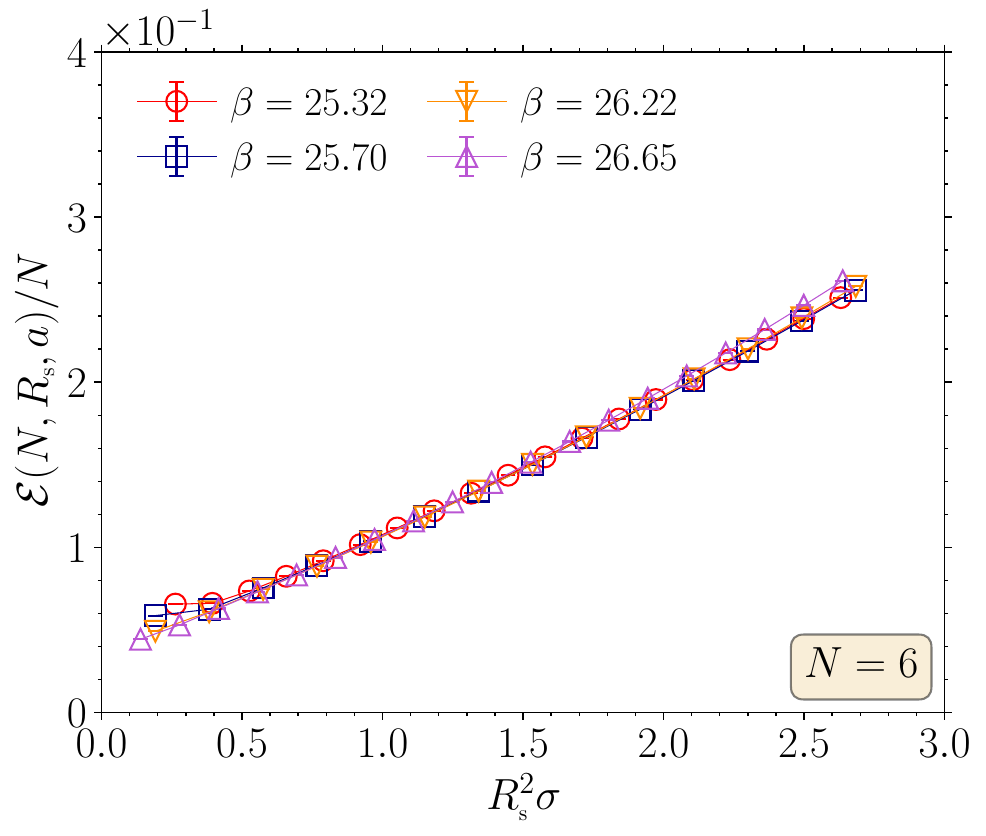}
\caption{Action density computed via the mean plaquette as a function of the smoothing radius, and for all explored values of $\beta$ and $N$.}
\label{fig:ene}
\end{figure}	

\begin{figure}[!t]	
\includegraphics[scale=0.25]{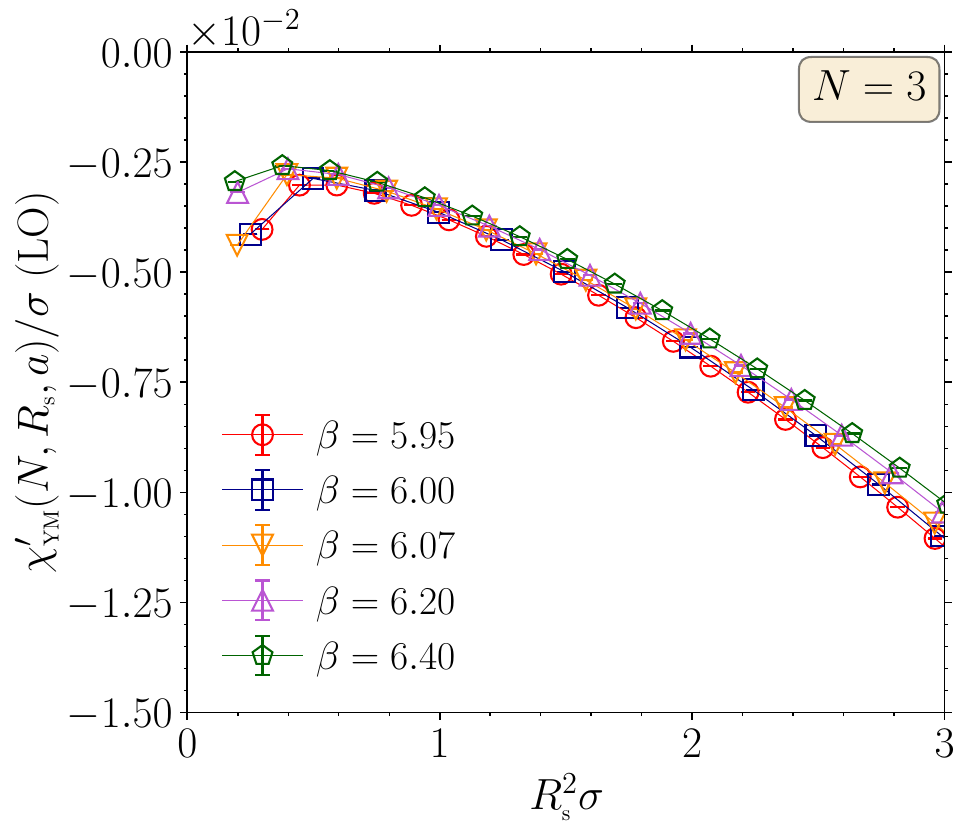}
\includegraphics[scale=0.25]{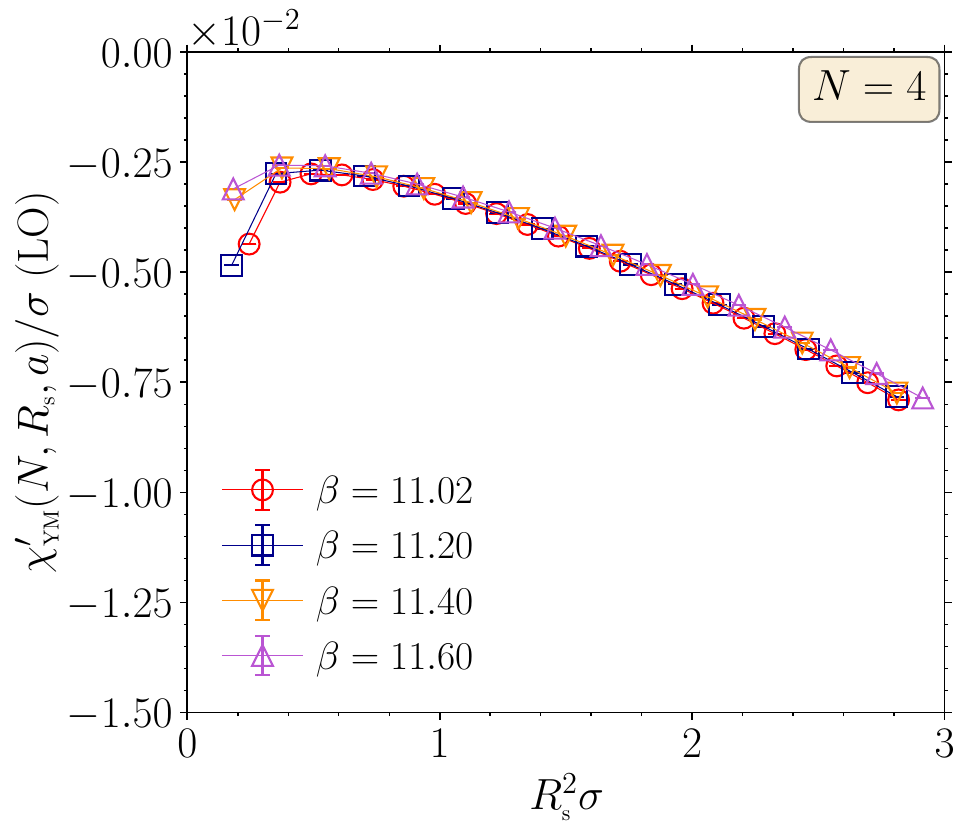}
\includegraphics[scale=0.25]{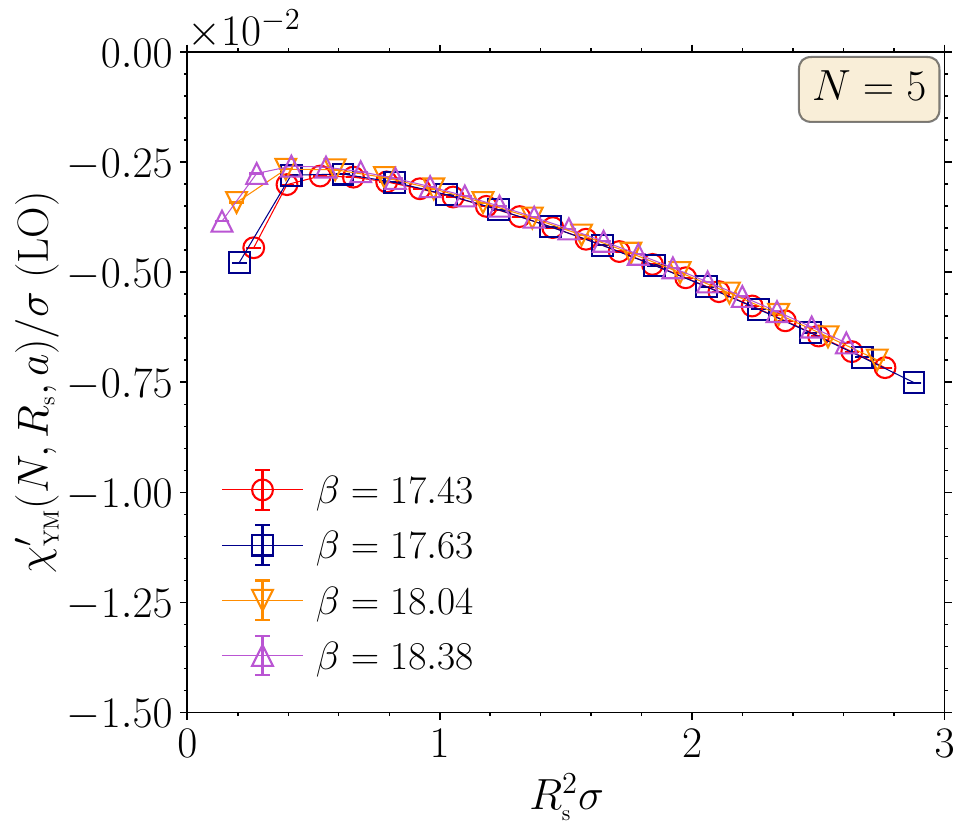}
\includegraphics[scale=0.25]{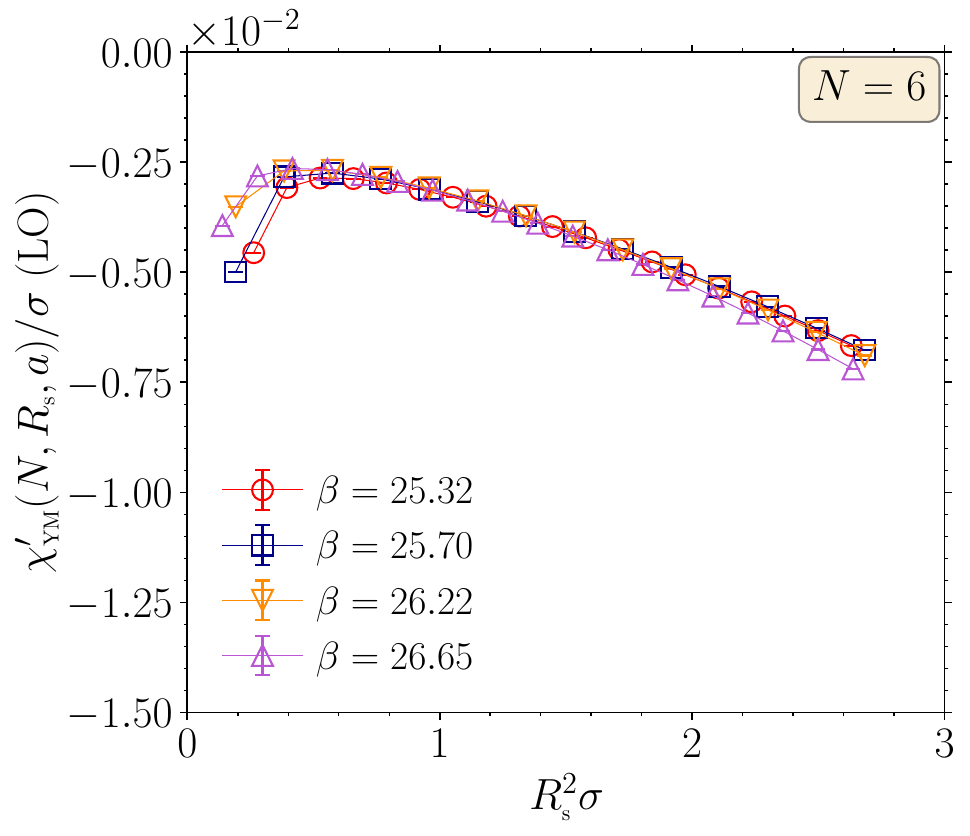}
\caption{Leading-order weak-coupling result for $\chip$, evaluated defining a lattice coupling via the action density, as a function of the smoothing radius, and for all explored values of $\beta$ and $N$.}
\label{fig:chip_LO}
\end{figure}

\begin{figure}[!t]
\centering
\includegraphics[scale=0.25]{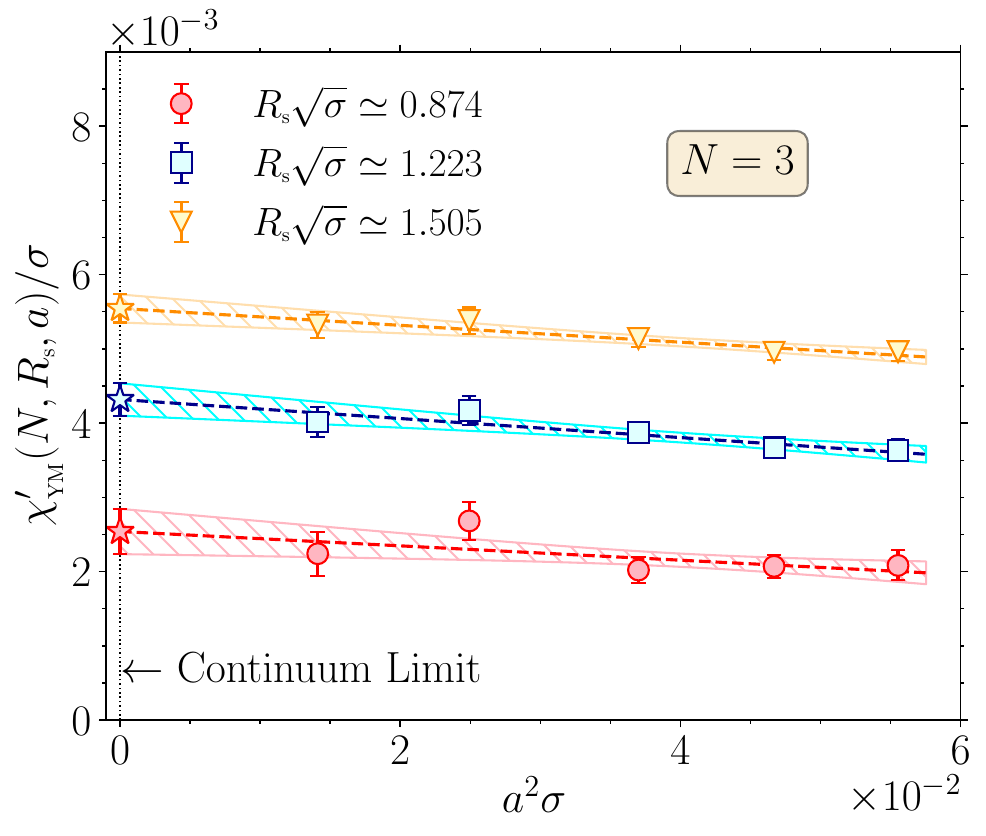}
\includegraphics[scale=0.25]{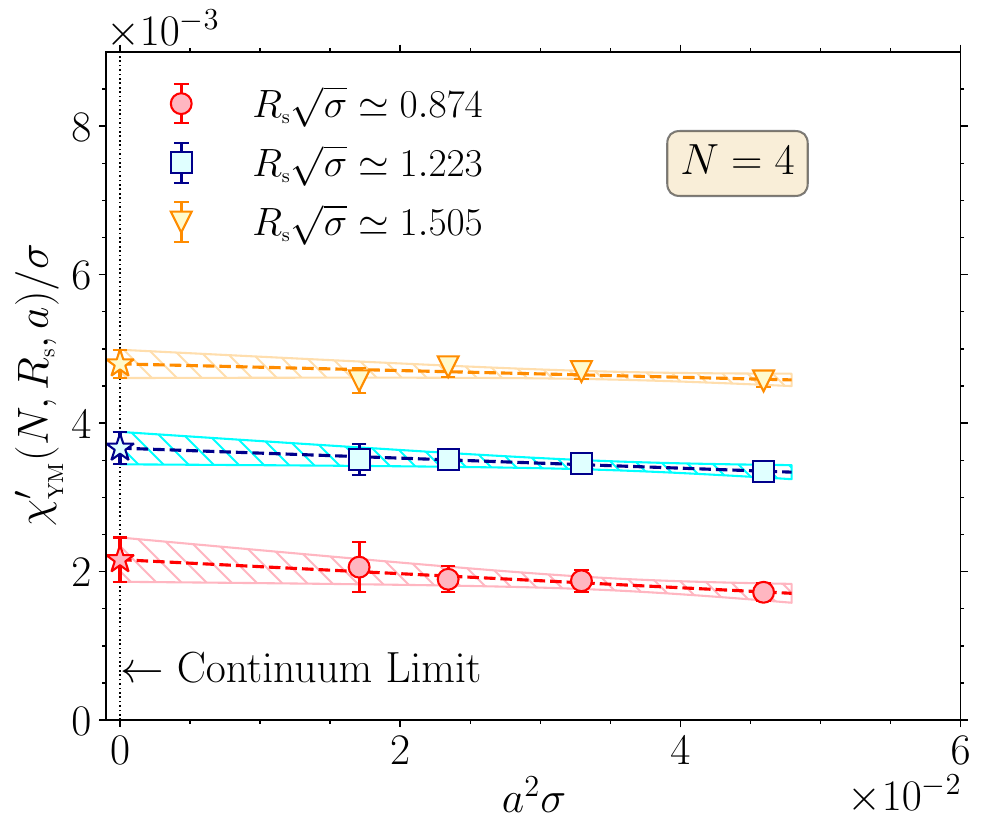}
\includegraphics[scale=0.25]{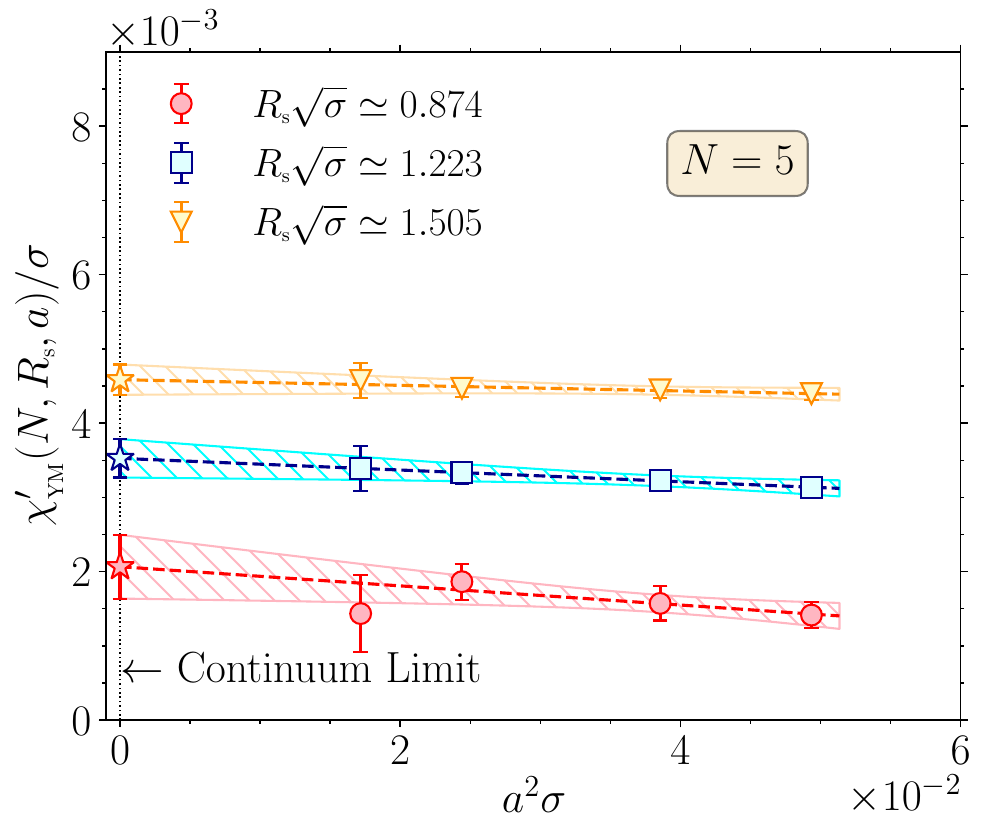}
\includegraphics[scale=0.25]{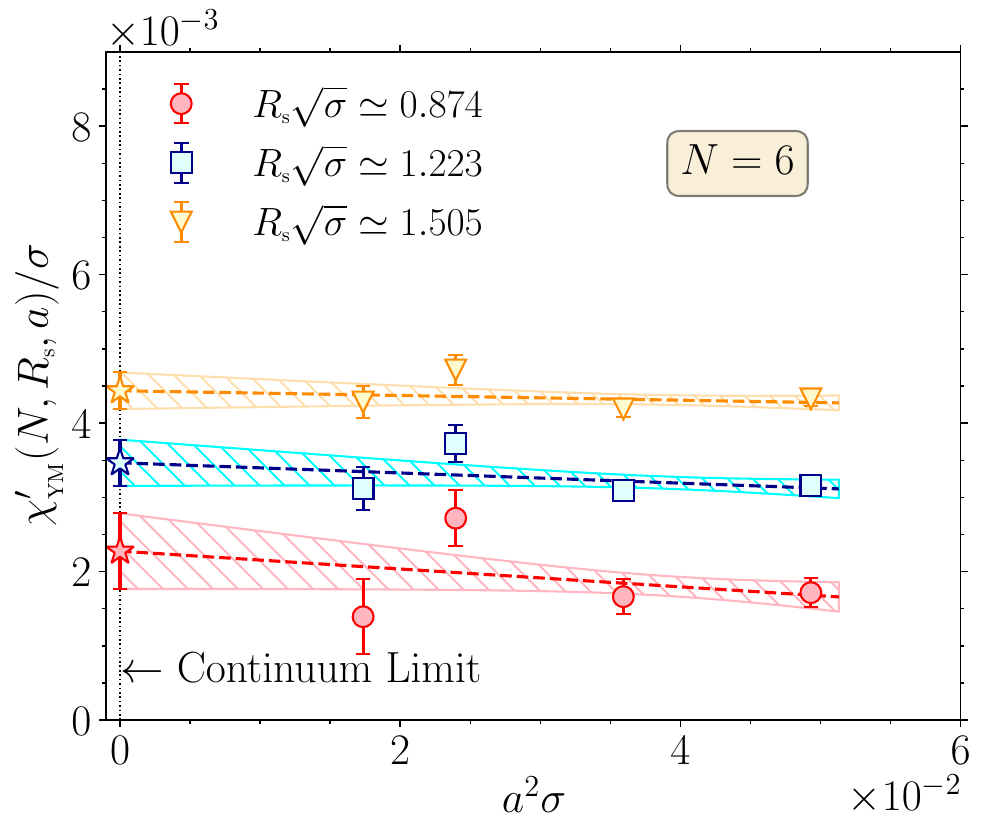}
\caption{Continuum limits of the unsubtracted $\chip_{_\YM}(N,\Rs,a)$ in units of the string tension $\sigma$ for $N=3,4,5,6$ and for three values of the smoothing radius $\Rs\sqrt{\sigma}\simeq 0.874,1.223,1.505$. Star points in $a^2\sigma=0$ represent the continuum extrapolations.}
\label{fig:cont_limit_nosub}
\end{figure}

On the lattice, one can adopt many possible discretizations of the action density. In this study I have considered the simplest one, namely, the one based on the average plaquette $P$ after cooling:
\bee
a^4 \braket{E}(N,\Rs,a) = 12N\left[1-\braket{P}(N,\Rs,a)\right], \qquad \qquad P \equiv \frac{1}{6NL^4}\sum_{x,\mu>\nu}\Re\Tr\left[U_{\mu\nu}(x)\right].
\eee
Based on the matching between cooling and gradient flow, $\Rs^2 = 8t = a^2 8 n_{\cool}/3$, one can write:
\bee
\mathcal{E}(N,\Rs,a) = \frac{1}{64}\left(\frac{\Rs}{a}\right)^4 a^4 \braket{E}(N,\Rs,a), \qquad\qquad \left(\frac{\Rs}{a}\right)^2 = \frac{8}{3}n_{\cool}.
\eee
Therefore, finally:
\bee\label{eq:final_expression_SLO}
\chip_{_\LO}(N,\Rs,a) = -\frac{8}{3\pi^2} \frac{\mathcal{E}^2(N,\Rs,a)}{(N^2-1)\Rs^2} \equiv \frac{S_{\LO}}{\Rs^2}.
\eee
Eq.~\eqref{eq:final_expression_SLO} reports the quantity $S_{\LO}/\Rs^2$ that is subtracted in Eq.~\eqref{eq:chip_subtraction_prescription} in the main text for the purpose of taking the $\Rs \to 0$ limit. In Figs.~\ref{fig:ene} and~\ref{fig:chip_LO}, lattice results for $\mathcal{E}$ and $\chip_{_\LO}$ as a function of $\Rs$, and for all explored values of $N$ and $\beta$, are shown. As it can be seen, lattice artifacts are small, meaning that the adopted subtraction prescription does not introduce further additional lattice artifacts. This can be also easily seen by comparing the continuum limits of the subtracted data in the main text, Fig.~\ref{fig:cont_limit}, with those of the unsubtracted ones, shown in this appendix in Fig.~\ref{fig:cont_limit_nosub}. It is also important to stress that both $\mathcal{E}/N$ and $\chip_{_\LO}$ show a very mild $N$ dependence, in agreement with theoretical expectations about their large-$N$ scaling.

\FloatBarrier

\stepcounter{count}
\section{A.\thecount~Estimation of systematic errors}

Statistical uncertainties throughout this work were computed by means of a binned bootstrap strategy, done at the level of the gauge configurations. This allows to take into account correlations among values of $\chip$ computed at different values of $\Rs$ and same $\beta,N$, as well as the correlation among $\chip$ and $\mathcal{E}$ data. All limits ($a \to 0$, $\Rs \to 0$, $N\to \infty$) were done bootstrap sample by bootstrap sample. Final central values and statistical errors are obtained as bootstrap averages.

In addition, several sources of systematic errors were taken into account, all described in detail below. On general grounds, for the purpose of estimating a systematic uncertainty between two different determinations obtained by means of two different strategies, yielding two results $O_1$ and $O_2$ with statistical errors $\Delta_1$ and $\Delta_2$, the systematic error was computed as follows:
\beq\label{eq:systematics}
\Delta_{\rm syst} = \vert O_1 - O_2 \vert \times \mathrm{erf}\left(\frac{\delta}{\sqrt{2}}\right), \qquad \delta = \frac{\vert O_1 - O_2 \vert}{\sqrt{\Delta_1^2 + \Delta_2^2}}
\eeq
The idea is to assign a systematic error that is proportional to the difference between the two outcomes $O_1$ and $O_2$, weighted by the probability that this difference is not due to a statistical fluctuation (erf is the error function).

\vs{+}
\noindent\textbf{Zero-smoothing-radius limit extrapolation}
\vs{+}

\begin{figure}[!b]
\centering
\includegraphics[scale=0.26]{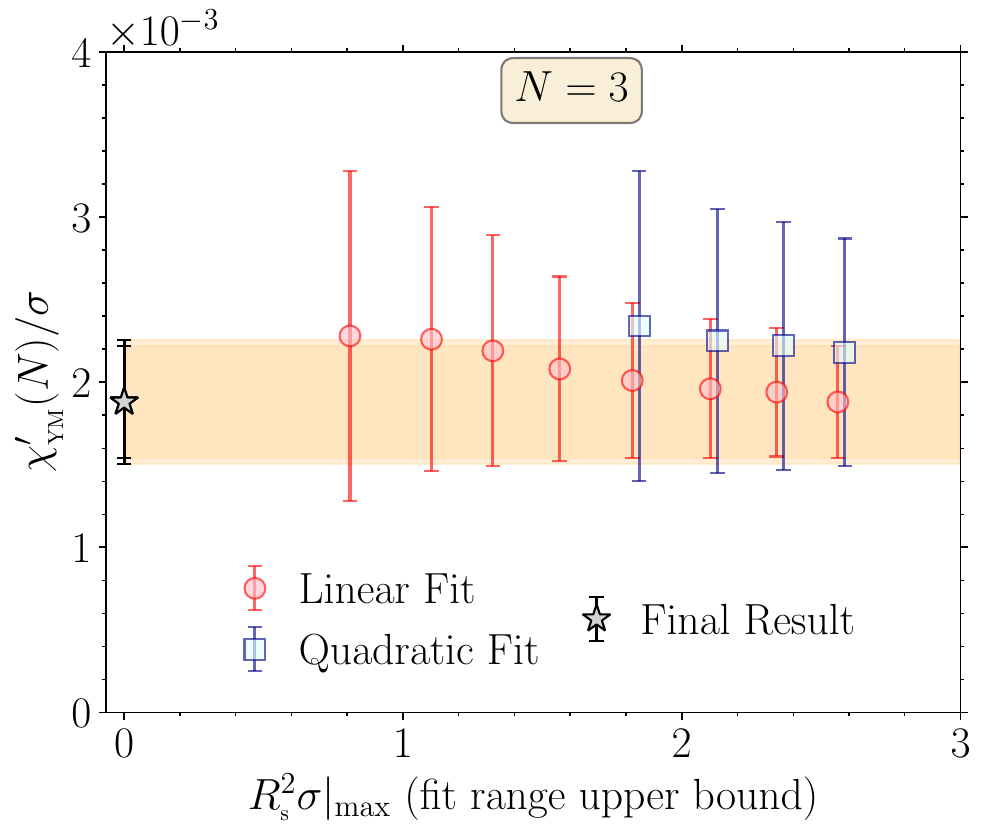}
\includegraphics[scale=0.26]{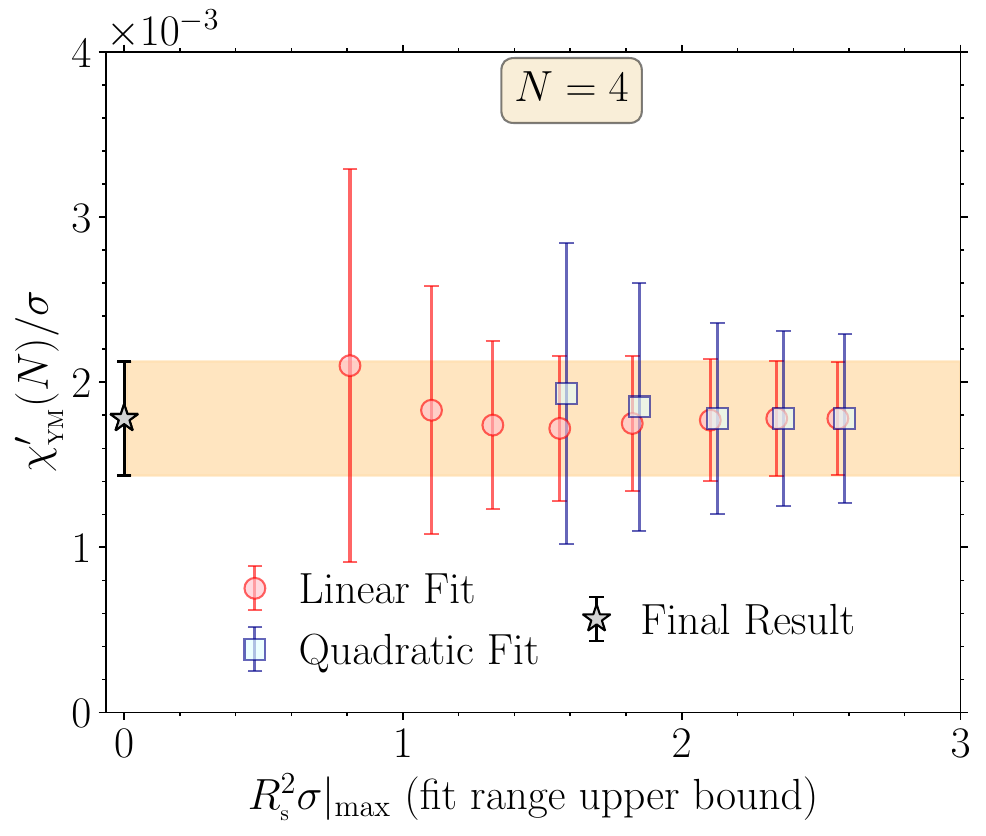}
\includegraphics[scale=0.26]{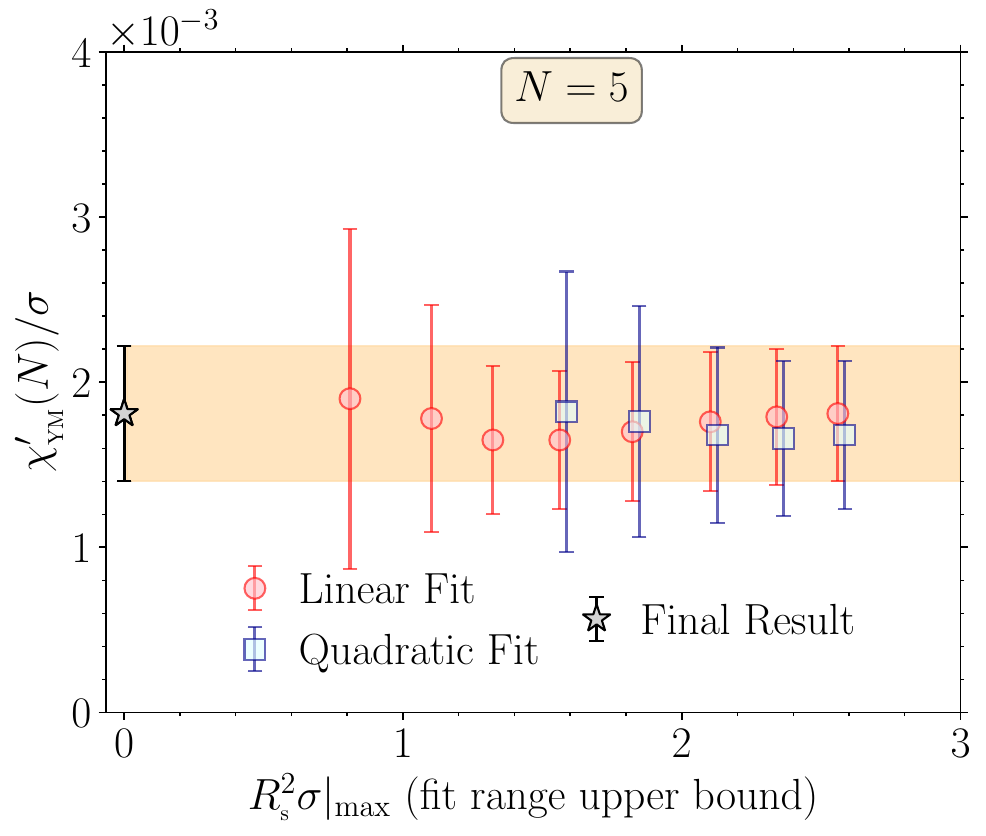}
\includegraphics[scale=0.26]{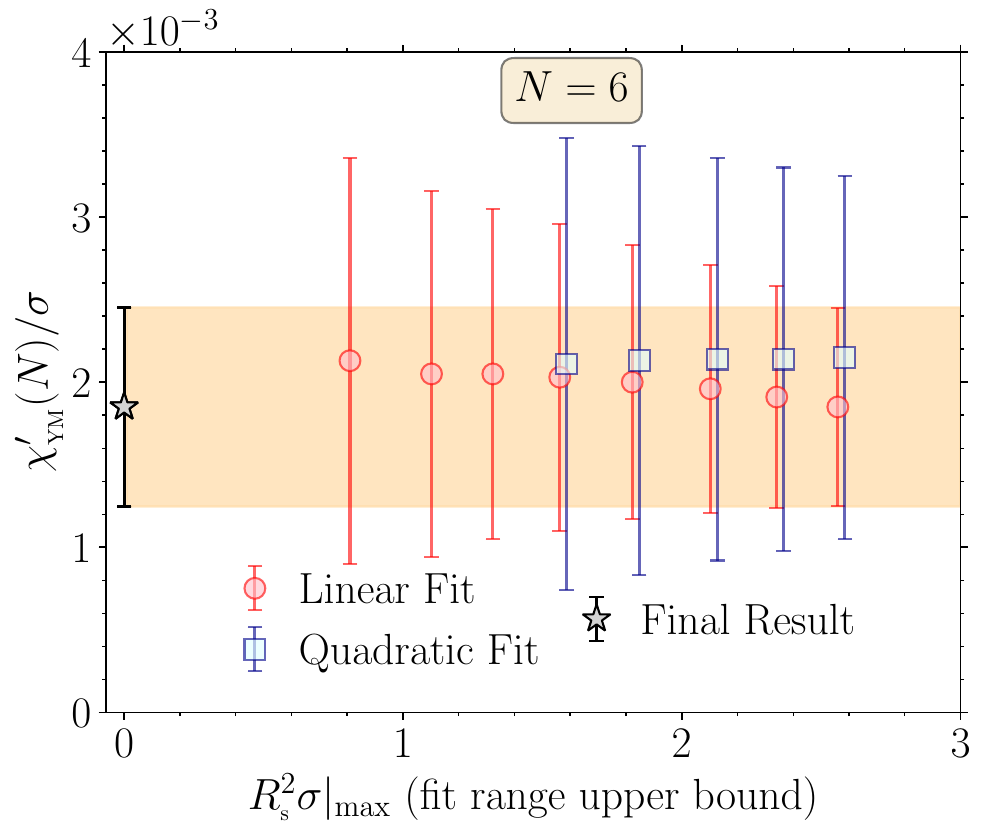}
\caption{Study of systematic effects on the $\Rs\to0$ extrapolation. Results obtained with linear and quadratic fits are shown as a function of the upper bound of the fit range $\Rs^2\sigma\vert_{\rm max}$.}
\label{fig:chip_Rs0_syst}
\end{figure}

Systematic uncertainties on the $\Rs \to 0$ extrapolations were assessed by performing several linear and quadratic fits of $\chipsub$ as a function of $\Rs^2\sigma$ varying the upper bound of the fit range (the lower bound was fixed to the smallest value of $\Rs$ available). The fit quality is in all cases very good, and both the $\Rs\to0$ and the linear slope turn out to be perfectly stable varying the upper bound of the fit range. Moreover, $\mathcal{O}(\Rs^4)$ terms turn out to be completely negligible in the explored range of smoothing radii: in all cases quadratic and linear extrapolations agree within errors, and the fitted coefficients of the $\Rs^4$ term are compatible with zero within errors in all cases.

The stability of the obtained $\Rs\to 0$ extrapolations is shown in Fig.~\ref{fig:chip_Rs0_syst}. The observed stability of the $\Rs \to 0$ extrapolation ultimately results in small systematic errors, all computed via Eq.~\eqref{eq:systematics}. The final systematic error is always taken to be the largest one observed among all best fits. The central value and statistical error of the final result were instead always determined from the outcome of the linear fit performed in the largest fit range.

\vs{+}
\noindent\textbf{Large-$N$ limit extrapolation}
\vs{+}

The systematic error on the large-$N$ limit of $\chip_{_\YM}$ is obtained considering two sources of systematic error: one coming from the systematic errors of the $\Rs\to 0$ extrapolated values of the susceptibility slope; and one associated to the difference between a constant best fit as a function of $N$ and a best fit including a further $\mathcal{O}(1/N^2)$ correction. Concerning the first source, systematic errors related to the $\Rs\to 0$ limits at each $N$ are propagated adding a uniformly-distributed random bias $\in [-\Delta_{\rm syst}(\Rs \to 0), \Delta_{\rm syst}(\Rs \to 0)]$ to $\chip_{_\YM}(N)$ on each bootstrap sample. Concerning the second source, for each bootstrap sample both the linear and the constant best fits are performed, and their difference gives the systematic error on the large-$N$ limit best fit function according to Eq.~\eqref{eq:systematics}. The latter turns out in the end to be completely negligible.

\vs{+}
\noindent\textbf{Physical-units conversion}
\vs{+}

In this work, the string tension $\sigma$ was used to express any dimensionful quantity in dimensionless units. Since Yang--Mills theories are not realized in Nature, assigning a value in MeV units to $\sqrt{\sigma}$ to express any other quantity in $\mathrm{MeV}$ units is an ambiguous procedure. In order to assign a systematic uncertainty to this ambiguity, I have considered several different prescriptions to pass $\sqrt{\sigma}$ to MeV units, summarized below. These are all constructed by fixing a certain quantity to the physical value it assumes for $N=3$, and then keeping this value fixed for all values of $N$.

The spread of the different results in MeV obtained from different prescriptions will be then used to define an averaged quantity. Apart from the total uncertainty (statistical + systematic) coming purely from the dimensionless ratio, the average quantity also has an additional systematic uncertainty related to the scale-setting ambiguity. This is computed as the root mean square deviation from the average of the outcomes of the different scale-setting prescriptions. All data obtained from different choices (including the final average) are reported in Tab.~\ref{tab:MeV_raw_data_chip_YM}.

\vs{0.5}
\noindent Scale setting prescriptions:
\begin{enumerate}
\item $\sqrt{\sigma}=445$ MeV~\cite{Bulava:2024jpj};
\item $\sqrt{8t_0} = 0.5$ fm~\cite{Ce:2015qha} $\implies$ $\sqrt{\sigma}=476$ MeV, using conversion factor $\sqrt{8t_0\sigma}=1.2068$~\cite{Bonanno:2025eeb};
\item $r_0$ = 0.5 fm~\cite{Giusti:2018cmp} $\implies$ $\sqrt{\sigma}=501$ MeV, using conversion factors $r_0/\sqrt{8t_0}=0.950$~\cite{Giusti:2018cmp} and $\sqrt{8t_0\sigma}$ at point 2.
\item $w_0 = 0.1757$ fm~\cite{BMW:2012hcm} $\implies$ $\sqrt{\sigma}=504$ MeV, using conversion factors $w_0^2/t_0=1.1058$~\cite{Bonanno:2025kfd} and $\sqrt{8t_0\sigma}$ at point 2.
\end{enumerate}

\begin{table}[!htb]
\centering
\begin{tabular}{|c|c|c|}
\hline
&&\\[-1em]
$N$ & $\sqrt{\sigma}$ [MeV] & $\sqrt{\chip_{_\YM}}$ [MeV]\\
&&\\[-1em]
\hline
\multirow{5}{*}{3} & 445   &   19.3(2.0)$_{\rm stat+syst}$ \\
 & 476   &   20.6(2.1)$_{\rm stat+syst}$ \\
 & 501   &   21.7(2.2)$_{\rm stat+syst}$ \\
 & 504   &   21.9(2.2)$_{\rm stat+syst}$ \\
\cline{2-3}
 &  average  & 20.9(2.1)$_{\rm stat+syst}$(1.0)$_{\rm scale}$ \\
\hline
\hline
\multirow{5}{*}{4} & 445   &   18.8(1.9)$_{\rm stat+syst}$ \\
 & 476   &   20.1(2.0)$_{\rm stat+syst}$ \\
 & 501   &   21.1(2.1)$_{\rm stat+syst}$ \\
 & 504   &   21.3(2.1)$_{\rm stat+syst}$ \\
\cline{2-3}
 &  average  & 20.3(2.0)$_{\rm stat+syst}$(1.0)$_{\rm scale}$ \\
\hline
\end{tabular}
\begin{tabular}{|c|c|c|}
\hline
&&\\[-1em]
$N$ & $\sqrt{\sigma}$ [MeV] & $\sqrt{\chip_{_\YM}}$ [MeV]\\
&&\\[-1em]
\hline
\multirow{5}{*}{5} & 445   &   18.9(2.1)$_{\rm stat+syst}$ \\
 & 476   &   20.3(2.3)$_{\rm stat+syst}$ \\
 & 501   &   21.3(2.4)$_{\rm stat+syst}$ \\
 & 504   &   21.4(2.4)$_{\rm stat+syst}$ \\
\cline{2-3}
 &  average  & 20.5(2.3)$_{\rm stat+syst}$(1.0)$_{\rm scale}$ \\
\hline
\hline
\multirow{5}{*}{6} & 445   &   19.1(3.1)$_{\rm stat+syst}$ \\
 & 476   &   20.5(3.3)$_{\rm stat+syst}$ \\
 & 501   &   21.6(3.5)$_{\rm stat+syst}$ \\
 & 504   &   21.7(3.5)$_{\rm stat+syst}$ \\
\cline{2-3}
 &  average  & 20.7(3.4)$_{\rm stat+syst}$(1.0)$_{\rm scale}$ \\
\hline
\end{tabular}
\begin{tabular}{|c|c|c|}
\hline
&&\\[-1em]
$N$ & $\sqrt{\sigma}$ [MeV] & $\sqrt{\chip_{_\YM}}$ [MeV]\\
&&\\[-1em]
\hline
\multirow{5}{*}{$\infty$} & 445   &   18.7(2.5)$_{\rm stat+syst}$ \\
& 476   &   20.0(2.7)$_{\rm stat+syst}$ \\
& 501   &   21.1(2.9)$_{\rm stat+syst}$ \\
& 504   &   21.2(2.9)$_{\rm stat+syst}$ \\
\cline{2-3}
&  average  & 20.3(2.8)$_{\rm stat+syst}$(1.0)$_{\rm scale}$ \\
\hline
\end{tabular}
\caption{Results for $\sqrt{\chip_{_\YM}(N)}$ in $\mathrm{MeV}$ units for several prescriptions to express $\sqrt{\sigma}$ in $\mathrm{MeV}$ units.}
\label{tab:MeV_raw_data_chip_YM}
\end{table}

\vs{+}
\newpage
\noindent\textbf{Chiral large-$N$ QCD vs real-world QCD}
\vs{+}

In the main text, a consistency test for the obtained results for $\chip_{_\QCD}$ in the large-$N$ and chiral limit is performed by evaluating $g_{\etap\sst{N}\sst{N}}$ via the Shore--Veneziano formula, and comparing it with experimental bounds. This procedure calls for an estimation of a systematic error associated to neglected finite-$N$/finite-$m$ corrections on $g_{\etap\sst{N}\sst{N}}$.

As explained in the main paper, this systematic uncertainty was computed [via Eq.~\eqref{eq:systematics}] comparing the outcome for the $\etap$-nucleon coupling obtained from the Shore--Veneziano formula using, as input, either the $N=\infty$ $m=0$ value of $\chip_{_\QCD}$, or a $N=3$ estimate with physical quarks masses obtained from the $\chi$PT formula and the following QCD inputs from FLAG~\cite{FlavourLatticeAveragingGroupFLAG:2021npn} (here $F$ is the result for the pion decay constant obtained in the $\Nf=3$ chiral limit):
\beq
F = 80.3(6.0) \text{ MeV}, \qquad\qquad z_{\rm ud} = m_{\rm u} / m_{\rm d} = 0.485(19), \qquad \qquad z_{\rm us} = m_{\rm u} / m_{\rm s} = 0.0246(12) .
\eeq
The results for $\chip_{_\QCD}$ in large-$N$ chiral QCD and in real-world QCD as a function of the different prescription for giving MeV units to $\sqrt{\sigma}$ are reported in Tab.~\ref{tab:MeV_raw_data_chip_QCD}. Each data is used to provide a result for $g_{\etap\sst{N}\sst{N}}$ from the Shore--Veneziano formula. These determinations are used to determine the final results for $\chip_{_\QCD}$ and $g_{\etap\sst{N}\sst{N}}$ reported in the main text, and to assign them statistical and systematic uncertainties.

\begin{table}[!t]
\centering
\begin{tabular}{|c|cc|cc|}
\hline
&&&&\\[-1em]
$\sqrt{\sigma}$ [MeV] & $\sqrt{\frac{3}{N}\vert\chip_{_\QCD}\vert}\Big\vert_{\substack{N\to\infty\\m\to 0}}$ [MeV] & $g_{\etap\sst{N}\sst{N}}$ & $\sqrt{\vert\chip_{_\QCD}\vert}\Big\vert_{\text{$\chi$PT}}$ [MeV] & $g_{\etap\sst{N}\sst{N}}$\\
&&&&\\[-1em]
\hline
445   & 39.7(1.1)$_{\sst{\text{$F$}}}$ & 2.60(48)$_{\sst{SV}}$ &   37.1(3.7)$_{\sst{\text{QCD\,inputs}}}$ &  2.78(54)$_{\sst{SV}}$ \\
476   & 42.5(1.1)$_{\sst{\text{$F$}}}$ & 2.43(45)$_{\sst{SV}}$ &   36.4(3.8)$_{\sst{\text{QCD\,inputs}}}$ &  2.84(51)$_{\sst{SV}}$ \\
501   & 44.7(1.2)$_{\sst{\text{$F$}}}$ & 2.31(42)$_{\sst{SV}}$ &   35.7(3.9)$_{\sst{\text{QCD\,inputs}}}$ &  2.89(49)$_{\sst{SV}}$ \\
504   & 45.0(1.2)$_{\sst{\text{$F$}}}$ & 2.29(42)$_{\sst{SV}}$ &   35.7(3.9)$_{\sst{\text{QCD\,inputs}}}$ &  2.90(49)$_{\sst{SV}}$ \\       
\hline
\end{tabular}
\caption{Results for $\sqrt{\vert\chip_{_\QCD}\vert}$ in $\mathrm{MeV}$ units and $g_{\etap\sst{N}\sst{N}}$ for several prescriptions to express $\sqrt{\sigma}$ in $\mathrm{MeV}$ units, and using results obtained either from $\chi\mathrm{PT}$ ($N=3$ and physical quark masses) or from the lattice ($N\to\infty$ and $m\to 0$).}
\label{tab:MeV_raw_data_chip_QCD}
\end{table}

\vs{-}
\stepcounter{count}
\section{A.\thecount~Cooling versus gradient flow}

\begin{figure}[!b]
\centering
\includegraphics[scale=0.49]{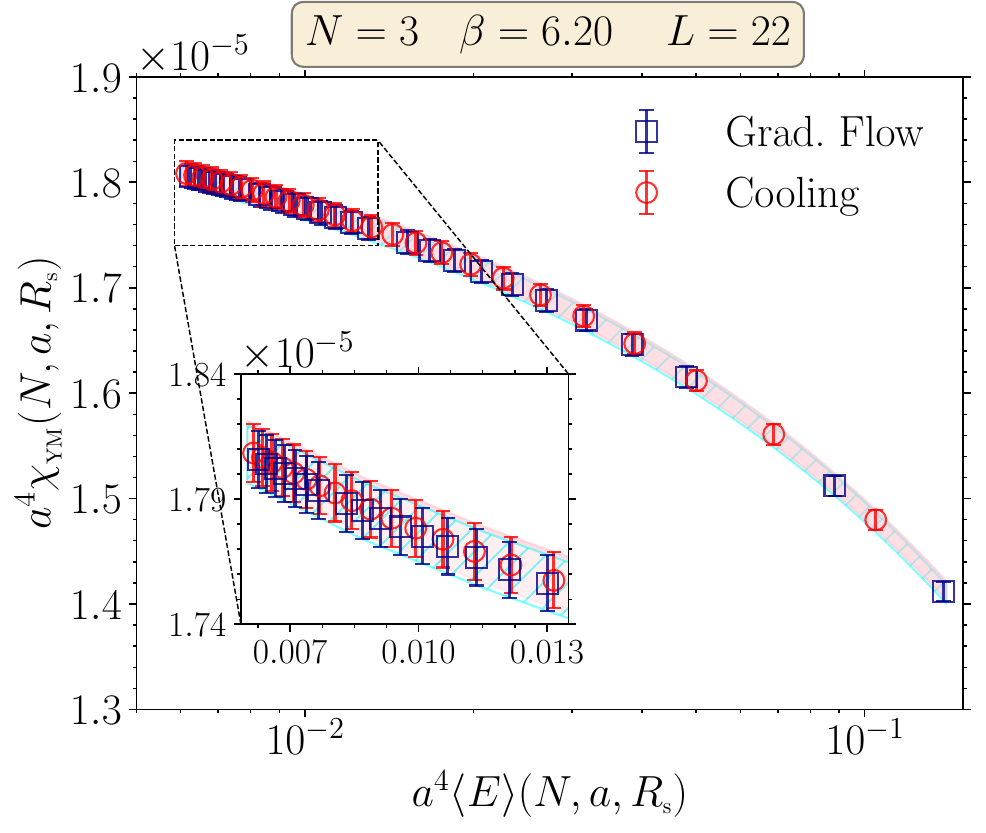}
\hfill
\includegraphics[scale=0.49]{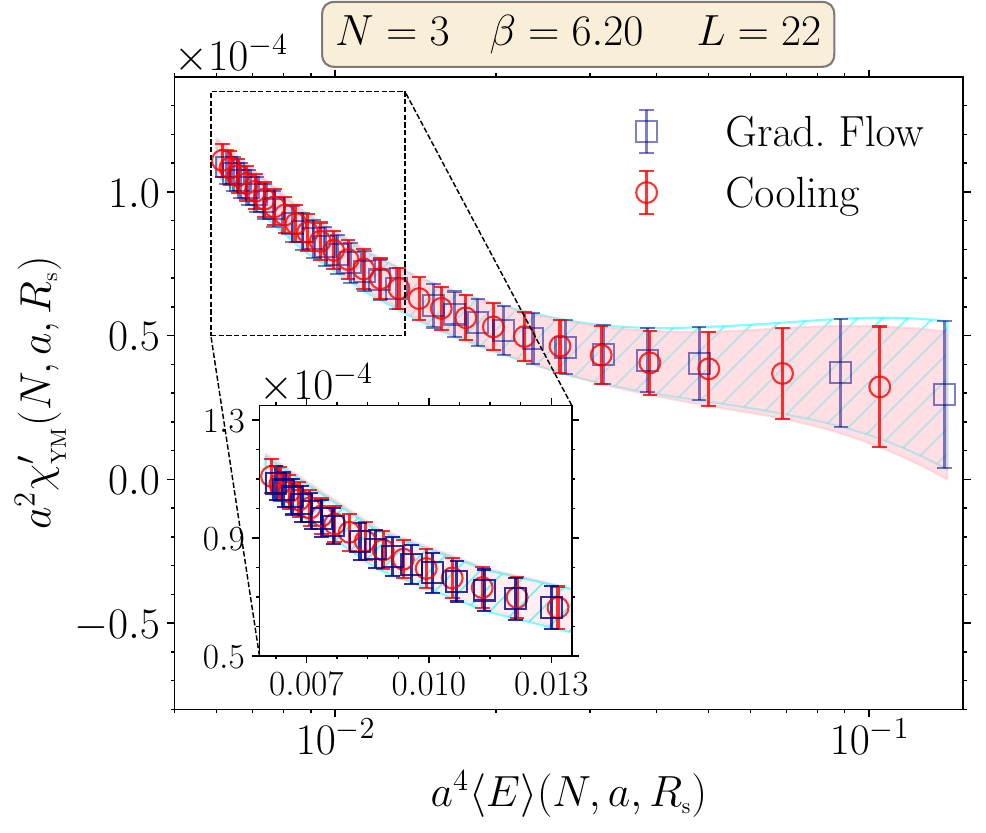}
\caption{Comparison of cooling and gradient flow topological quantities as a function of the action density, used as thermometer. Left panel: topological susceptibility. Right panel: topological susceptibility slope.}
\label{fig:cooling_vs_gradientflow}
\end{figure}

It was shown in Ref.~\cite{Bonati:2014tqa} that Wilson cooling and Wilson gradient flow give perfectly agreeing results for the topological susceptibility and the action density once the number of cooling steps and the flow time are appropriately matched to yield the same smoothing radius: $t/a^2=n_\cool/3$. Later on, this equivalence was extended to other action discretizations and other smoothing methods~\cite{Alexandrou:2015yba,Alexandrou:2017hqw}. The purpose of this appendix is to show that such equivalence holds for the topological susceptibility slope too. To this end, an additional simulation for one lattice spacing of $N=3$ was performed, with the purpose of computing the action density, the topological charge and $\chip_{_\YM}$ after the Wilson flow, integrated via an adaptive-step third-order Runge--Kutta scheme.

The comparison between Wilson cooling and Wilson gradient flow is shown in Fig.~\ref{fig:cooling_vs_gradientflow}, where $\chip_{_\YM}$ (right panel) is reported as a function of the action density, used as ``thermometer''. As a consistency check with~\cite{Bonati:2014tqa}, the comparison of $\chi$ is also shown (left panel). Results obtained with the two smoothing methods perfectly agree, confirming their equivalence for the topological susceptibility slope too. At fixed smoothing radius range and measurement frequency, the computational cost of cooling is found to be almost an order of magnitude smaller than gradient flow. In light of their equivalence, this justifies the adoption of cooling to perform this study, given both the very large statistics necessary to compute $\chip$, and the fairly large number of lattice ensembles here considered.

\stepcounter{count}
\section{A.\thecount~Raw numerical data}

All raw data for $\chip_{_\YM}$ in lattice units and for the plaquette $P$ for all values of $N$, $\beta$ and $n_\cool$ obtained in this study are collected, respectively, in Tab.~\ref{tab:rawdata_chip} and Tab.~\ref{tab:rawdata_ene}.

\vspace*{2\baselineskip}
\begin{table}[!htb]
\scriptsize
\begin{center}
\begin{tabular}{|c|cc|cc|cc|cc|cc|}
\hline
\multicolumn{11}{|c|}{}\\
\multicolumn{11}{|c|}{{\small$n_\cool\quad$ versus $\quad10^4 \times a^2 \chip_{_\YM}$}}\\
\multicolumn{11}{|c|}{}\\
\hline
&&&&&&&&&&\\
$N$ &\multicolumn{2}{c|}{$\beta=5.95$} &\multicolumn{2}{c|}{$\beta=6.00$}&\multicolumn{2}{c|}{$\beta=6.07$}&\multicolumn{2}{c|}{$\beta=6.20$}&\multicolumn{2}{c|}{$\beta=6.40$}\\
&&&&&&&&&&\\
\hline
\multirow{18}{*}{$\,\,$ 3 $\,$ }
& 2  & 0.16(20)   &  2  &  0.13(15)   &  2  &  -0.04(17)  &  3  &  0.18(17)  &  5  &  -0.07(17) \\
& 3  &  0.61(16)  &  4  &  0.627(92)  &  4  &  0.261(97)  &  6  &  0.454(93) &  10 &  0.087(75) \\
& 4  &  0.89(13)  &  6  &  0.948(73)  &  6  &  0.534(74)  &  9  &  0.577(71) &  15 &  0.211(52) \\
& 5  &  1.12(11)  &  8  &  1.227(63)  &  8  &  0.778(64)  &  12 &  0.688(62) &  20 &  0.311(43) \\
& 6  &  1.33(10)  &  10 &  1.479(56)  &  10 &  0.992(57)  &  15 &  0.797(57) &  25 &  0.392(37) \\
& 7  &  1.512(96) &  12 &  1.710(52)  &  12 &  1.176(53)  &  18 &  0.898(53) &  30 &  0.452(32) \\
& 8  &  1.683(90) &  14 &  1.918(48)  &  14 &  1.344(49)  &  21 &  0.995(51) &  35 &  0.514(30) \\
& 9  &  1.848(85) &  16 &  2.115(46)  &  16 &  1.491(46)  &  24 &  1.082(49) &  40 &  0.569(28) \\
& 10 &  2.003(81) &  18 &  2.296(43)  &  18 &  1.620(44)  &  27 &  1.165(47) &  45 &  0.621(27) \\
& 11 &  2.152(78) &  20 &  2.463(41)  &  20 &  1.743(43)  &  30 &  1.242(45) &  50 &  0.666(26) \\
& 12 &  2.301(75) &  22 &  2.619(40)  &  22 &  1.852(41)  &  33 &  1.315(44) &  55 &  0.709(25) \\
& 13 &  2.448(72) &     &             &  24 &  1.955(40)  &  36 &  1.385(43) &  60 &  0.749(24) \\
& 14 &  2.587(70) &     &             &  26 &  2.054(38)  &  39 &  1.451(42) &  65 &  0.787(24) \\
& 15 &  2.717(68) &     &             &  28 &  2.153(37)  &  42 &  1.515(41) &  70 &  0.823(23) \\
& 16 &  2.844(66) &     &             &     &             &     &            &  75 &  0.857(23) \\
& 17 &  2.966(64) &&&&&&&&\\
& 18 &  3.083(62) &&&&&&&&\\
& 19 &  3.194(61) &&&&&&&&\\
\hline
\end{tabular}
\hspace*{2.5\baselineskip}
\begin{tabular}{|c|cc|cc|cc|cc|}
\hline
\multicolumn{9}{|c|}{}\\
\multicolumn{9}{|c|}{{\small$n_\cool\quad$ versus $\quad10^4 \times a^2 \chip_{_\YM}$}}\\
\multicolumn{9}{|c|}{}\\
\hline
&&&&&&&&\\
$N$ & \multicolumn{2}{c|}{$\beta=11.02$} &\multicolumn{2}{c|}{$\beta=11.20$}&\multicolumn{2}{c|}{$\beta=11.40$}&\multicolumn{2}{c|}{$\beta=11.60$}\\
&&&&&&&&\\
\hline
\multirow{18}{*}{$\,\,$ 4 $\,$ }
&  2  &  -0.15(13)  &  2  &  -0.13(29)  &  3  &  -0.12(16)  &  4  &  -0.48(28) \\
&  3  &  0.150(93)  &  4  &  0.163(94)  &  6  &  0.167(74)  &  8  &  -0.02(12) \\
&  4  &  0.383(74)  &  6  &  0.393(65)  &  9  &  0.309(50)  &  12 &  0.205(77) \\
&  5  &  0.580(63)  &  8  &  0.564(52)  &  12 &  0.435(42)  &  16 &  0.332(59) \\
&  6  &  0.752(57)  &  10 &  0.714(45)  &  15 &  0.548(37)  &  20 &  0.417(50) \\
&  7  &  0.908(52)  &  12 &  0.848(41)  &  18 &  0.649(35)  &  24 &  0.483(44) \\
&  8  &  1.049(49)  &  14 &  0.970(38)  &  21 &  0.741(34)  &  28 &  0.539(40) \\
&  9  &  1.177(47)  &  16 &  1.083(36)  &  24 &  0.825(33)  &  32 &  0.589(37) \\
&  10 &  1.297(45)  &  18 &  1.189(34)  &  27 &  0.903(32)  &  36 &  0.636(34) \\
&  11 &  1.410(43)  &  20 &  1.288(33)  &  30 &  0.975(32)  &  40 &  0.681(32) \\
&  12 &  1.517(42)  &  22 &  1.381(32)  &  33 &  1.044(31)  &  44 &  0.723(31) \\
&  13 &  1.618(41)  &  24 &  1.470(31)  &  36 &  1.109(31)  &  48 &  0.763(29) \\
&  14 &  1.714(40)  &  26 &  1.554(30)  &  39 &  1.171(30)  &  52 &  0.801(28) \\
&  15 &  1.806(39)  &  28 &  1.635(29)  &  42 &  1.230(30)  &  56 &  0.838(27) \\
&  16 &  1.894(38)  &  30 &  1.712(29)  &  45 &  1.286(30)  &  60 &  0.873(27) \\
&  17 &  1.978(38)  &  32 &  1.786(28)  &     &             &  64 &  0.906(26) \\
&  18 &  2.059(37)  &&&&&&\\
&  19 &  2.137(36)  &&&&&&\\
&  20 &  2.213(36)  &&&&&&\\
&  21 &  2.286(36)  &&&&&&\\
\hline
\end{tabular}
\begin{tabular}{|c|cc|cc|cc|cc|}
\hline
\multicolumn{9}{|c|}{}\\
\multicolumn{9}{|c|}{{\small$n_\cool\quad$ versus $\quad10^4 \times a^2 \chip_{_\YM}$}}\\
\multicolumn{9}{|c|}{}\\
\hline
&&&&&&&&\\
$N$ & \multicolumn{2}{c|}{$\beta=17.43$} &\multicolumn{2}{c|}{$\beta=17.63$}&\multicolumn{2}{c|}{$\beta=18.04$}&\multicolumn{2}{c|}{$\beta=18.375$}\\
&&&&&&&&\\
\hline
\multirow{18}{*}{$\,\,$ 5 $\,$ }
& 2  &  -0.07(23)  &  2  &  -0.24(32)  &  3  &  0.24(26)   &  3  &  -0.772(671) \\
& 3  &  0.054(16)  &  4  &  0.12(16)   &  6  &  0.26(12)   &  6  &  -0.489(274) \\
& 4  &  0.31(12)   &  6  &  0.43(11)   &  9  &  0.353(75)  &  9  &  -0.188(167) \\
& 5  &  0.536(97)  &  8  &  0.670(83)  &  12 &  0.465(58)  &  12 &  0.033(122) \\
& 6  &  0.737(84)  &  10 &  0.871(69)  &  15 &  0.573(48)  &  15 &  0.183(98) \\
& 7  &  0.917(74)  &  12 &  1.047(61)  &  18 &  0.670(42)  &  18 &  0.290(83) \\
& 8  &  1.081(67)  &  14 &  1.203(55)  &  21 &  0.759(38)  &  21 &  0.371(73) \\
& 9  &  1.231(62)  &  16 &  1.346(50)  &  24 &  0.841(35)  &  24 &  0.437(65) \\
& 10 &  1.372(58)  &  18 &  1.477(47)  &  27 &  0.916(32)  &  27 &  0.493(59) \\
& 11 &  1.503(54)  &  20 &  1.599(44)  &  30 &  0.987(31)  &  30 &  0.542(55) \\
& 12 &  1.626(52)  &  22 &  1.712(41)  &  33 &  1.053(29)  &  33 &  0.587(51) \\
& 13 &  1.742(49)  &  24 &  1.820(39)  &  36 &  1.115(28)  &  36 &  0.628(48) \\
& 14 &  1.851(47)  &  26 &  1.921(38)  &  39 &  1.173(27)  &  39 &  0.667(46) \\
& 15 &  1.956(45)  &  28 &  2.016(36)  &  42 &  1.229(26)  &  42 &  0.703(43) \\
& 16 &  2.055(44)  &     &             &     &             &  45 &  0.737(42) \\
& 17 &  2.150(42)  &     &             &     &             &  48 &  0.770(40) \\
& 18 &  2.241(41)  &     &             &     &             &  51 &  0.801(39) \\
& 19 &  2.328(40)  &     &             &     &             &  54 &  0.831(38) \\
& 20 &  2.411(39)  &     &             &     &             &  57 &  0.860(36) \\
& 21 &  2.492(38)  &&&&&&\\
\hline
\end{tabular}
\hspace*{7.65\baselineskip}
\begin{tabular}{|c|cc|cc|cc|cc|}
\hline
\multicolumn{9}{|c|}{}\\
\multicolumn{9}{|c|}{{\small$n_\cool\quad$ versus $\quad10^4 \times a^2 \chip_{_\YM}$}}\\
\multicolumn{9}{|c|}{}\\
\hline
&&&&&&&&\\
$N$ & \multicolumn{2}{c|}{$\beta=25.32$} &\multicolumn{2}{c|}{$\beta=25.70$}&\multicolumn{2}{c|}{$\beta=26.22$}&\multicolumn{2}{c|}{$\beta=26.65$}\\
&&&&&&&&\\
\hline
\multirow{18}{*}{$\,\,$ 6 $\,$ }
&  2  &  0.24(26)   &  2  &  0.12(35)   &  3  &  0.50(46)   &  3  &  1.07(66) \\
&  3  &  0.33(17)   &  4  &  0.18(16)   &  6  &  0.59(19)   &  6  &  0.13(27) \\
&  4  &  0.52(13)   &  6  &  0.41(11)   &  9  &  0.61(12)   &  9  &  0.05(16) \\
&  5  &  0.71(11)   &  8  &  0.600(83)  &  12 &  0.653(90)  &  12 &  0.12(12) \\
&  6  &  0.877(93)  &  10 &  0.758(69)  &  15 &  0.709(76)  &  15 &  0.201(95) \\
&  7  &  1.025(82)  &  12 &  0.895(60)  &  18 &  0.774(68)  &  18 &  0.278(80) \\
&  8  &  1.160(75)  &  14 &  1.018(53)  &  21 &  0.840(62)  &  21 &  0.346(70) \\
&  9  &  1.286(69)  &  16 &  1.131(49)  &  24 &  0.905(58)  &  24 &  0.406(63) \\
&  10 &  1.405(65)  &  18 &  1.237(45)  &  27 &  0.968(55)  &  27 &  0.459(57) \\
&  11 &  1.519(61)  &  20 &  1.337(42)  &  30 &  1.028(52)  &  30 &  0.507(53) \\
&  12 &  1.628(58)  &  22 &  1.432(40)  &  33 &  1.085(50)  &  33 &  0.551(49) \\
&  13 &  1.732(56)  &  24 &  1.521(38)  &  36 &  1.140(49)  &  36 &  0.592(46) \\
&  14 &  1.833(54)  &  26 &  1.606(36)  &  39 &  1.191(47)  &  39 &  0.630(44) \\
&  15 &  1.929(52)  &  28 &  1.687(35)  &  42 &  1.241(46)  &  42 &  0.666(42) \\
&  16 &  2.023(50)  &     &             &     &             &  45 &  0.700(40) \\
&  17 &  2.113(49)  &     &             &     &             &  48 &  0.733(38) \\
&  18 &  2.200(47)  &     &             &     &             &  51 &  0.764(37) \\
&  19 &  2.284(46)  &     &             &     &             &  54 &  0.794(36) \\
&  20 &  2.366(45)  &     &             &     &             &  57 &  0.823(35) \\
\hline
\end{tabular}
\end{center}
\caption{Each column reports $10^4 \times a^2\chip_{_\YM}$ as a function of $n_\cool$ for a given $\beta$ and $N$.}
\label{tab:rawdata_chip}
\end{table}

\newpage

\begin{table}[!t]
\tiny
\centering
\begin{tabular}{|c|cc|cc|cc|cc|cc|}
\hline
\multicolumn{11}{|c|}{}\\
\multicolumn{11}{|c|}{{\small$n_\cool\quad$ versus $\quad P$}}\\
\multicolumn{11}{|c|}{}\\
\hline
&&&&&&&&&&\\
$N$ &\multicolumn{2}{c|}{$\beta=5.95$} &\multicolumn{2}{c|}{$\beta=6.00$}&\multicolumn{2}{c|}{$\beta=6.07$}&\multicolumn{2}{c|}{$\beta=6.20$}&\multicolumn{2}{c|}{$\beta=6.40$}\\
&&&&&&&&&&\\
\hline
\multirow{18}{*}{$\,\,$ 3 $\,$ }
& 2  &  0.988249347(165) &  2  &  0.989078179(144) &  2  &  0.989989218(100) &  3  &  0.996187135(58)  &  5  &  0.9987193326(158) \\
& 3  &  0.994457919(136) &  4  &  0.996783583(108) &  4  &  0.997180701(77)  &  6  &  0.998770700(49)  &  10 &  0.9995760110(121) \\
& 4  &  0.996401767(122) &  6  &  0.998167725(96)  &  6  &  0.998442962(69)  &  9  &  0.999314591(46)  &  15 &  0.9997641852(107) \\
& 5  &  0.997346584(114) &  8  &  0.998719537(89)  &  8  &  0.998938038(65)  &  12 &  0.999531260(44)  &  20 &  0.9998394876(100) \\
& 6  &  0.997897454(108) &  10 &  0.999008315(84)  &  10 &  0.999193280(62)  &  15 &  0.999643587(43)  &  25 &  0.9998785255(95) \\
& 7  &  0.998254370(103) &  12 &  0.999183983(80)  &  12 &  0.999346399(60)  &  18 &  0.999711152(42)  &  30 &  0.9999019715(92) \\
& 8  &  0.998502772(99)  &  14 &  0.999301666(77)  &  14 &  0.999447635(58)  &  21 &  0.999755887(41)  &  35 &  0.9999174625(89) \\
& 9  &  0.998684911(96)  &  16 &  0.999385979(74)  &  16 &  0.999519270(56)  &  24 &  0.999787567(40)  &  40 &  0.9999284061(87) \\
& 10 &  0.998823889(93)  &  18 &  0.999449445(71)  &  18 &  0.999572559(54)  &  27 &  0.999811143(39)  &  45 &  0.9999365302(85) \\
& 11 &  0.998933314(90)  &  20 &  0.999499061(69)  &  20 &  0.999613755(53)  &  30 &  0.999829373(38)  &  50 &  0.9999427956(83) \\
& 12 &  0.999021684(87)  &  22 &  0.999539021(67)  &  22 &  0.999646585(52)  &  33 &  0.999843901(37)  &  55 &  0.9999477762(82) \\
& 13 &  0.999094561(85)  &     &                   &  24 &  0.999673399(50)  &  36 &  0.999855766(37)  &  60 &  0.9999518335(81) \\
& 14 &  0.999155724(83)  &     &                   &  26 &  0.999695746(49)  &  39 &  0.999865655(36)  &  65 &  0.9999552064(80) \\
& 15 &  0.999207829(81)  &     &                   &  28 &  0.999714692(48)  &  42 &  0.999874037(35)  &  70 &  0.9999580585(78) \\
& 16 &  0.999252789(80)  &     &                   &     &                   &     &                   &  75 &  0.9999605054(77) \\
& 17 &  0.999292019(78)  &&&&&&&&\\
& 18 &  0.999326584(76)  &&&&&&&&\\
& 19 &  0.999357300(75)  &&&&&&&&\\
\hline
\end{tabular}
\begin{tabular}{|c|cc|cc|cc|cc|}
\hline
\multicolumn{9}{|c|}{}\\
\multicolumn{9}{|c|}{{\small$n_\cool\quad$ versus $P$}}\\
\multicolumn{9}{|c|}{}\\
\hline
&&&&&&&&\\
$N$ & \multicolumn{2}{c|}{$\beta=11.02$} &\multicolumn{2}{c|}{$\beta=11.20$}&\multicolumn{2}{c|}{$\beta=11.40$}&\multicolumn{2}{c|}{$\beta=11.60$}\\
&&&&&&&&\\
\hline
\multirow{18}{*}{$\,\,$ 4 $\,$ }
& 2 &  0.988579831(94)  &  2 &  0.989809367(58)  &  3 &  0.996107658(26)  &  4 &  0.997921953(26) \\
& 3 &  0.994881890(77)  &  4 &  0.997278518(43)  &  6 &  0.998780494(21)  &  8 &  0.999332041(18) \\
& 4 &  0.996783149(70)  &  6 &  0.998540390(38)  &  9 &  0.999334028(19)  &  12 &  0.999635263(16) \\
& 5 &  0.997688482(66)  &  8 &  0.999028637(36)  &  12 &  0.999552745(18)  &  16 &  0.999755531(15) \\
& 6 &  0.998208937(64)  &  10 &  0.999277585(34)  &  15 &  0.999665313(17)  &  20 &  0.999817466(14) \\
& 7 &  0.998542061(61)  &  12 &  0.999425400(33)  &  18 &  0.999732561(17)  &  24 &  0.999854449(13) \\
& 8 &  0.998771279(60)  &  14 &  0.999522179(32)  &  21 &  0.999776794(16)  &  28 &  0.999878753(13) \\
& 9 &  0.998937540(58)  &  16 &  0.999590019(31)  &  24 &  0.999807918(16)  &  32 &  0.999895835(12) \\
& 10 &  0.999063092(57)  &  18 &  0.999640029(30)  &  27 &  0.999830937(15)  &  36 &  0.999908454(12) \\
& 11 &  0.999160965(56)  &  20 &  0.999678351(29)  &  30 &  0.999848628(15)  &  40 &  0.999918140(12) \\
& 12 &  0.999239249(55)  &  22 &  0.999708630(28)  &  33 &  0.999862643(15)  &  44 &  0.999925803(11) \\
& 13 &  0.999303210(54)  &  24 &  0.999733156(28)  &  36 &  0.999874022(14)  &  48 &  0.999932018(11) \\
& 14 &  0.999356411(53)  &  26 &  0.999753434(27)  &  39 &  0.999883452(14)  &  52 &  0.999937162(11) \\
& 15 &  0.999401337(52)  &  28 &  0.999770491(27)  &  42 &  0.999891401(14)  &  56 &  0.999941493(11) \\
& 16 &  0.999439777(51)  &  30 &  0.999785051(26)  &  45 &  0.999898199(14)  &  60 &  0.999945193(10) \\
& 17 & 0.999473042(50)   &  32 &  0.999797636(26)  &     &                   &  64 & 0.999948393(10) \\
& 18 & 0.999502118(50)  &&&&&&\\
& 19 & 0.999527757(49)  &&&&&&\\
& 20 & 0.999550545(48)  &&&&&&\\
& 21 & 0.999570940(48)  &&&&&&\\
\hline
\end{tabular}
\begin{tabular}{|c|cc|cc|cc|cc|}
\hline
\multicolumn{9}{|c|}{}\\
\multicolumn{9}{|c|}{{\small$n_\cool\quad$ versus $P$}}\\
\multicolumn{9}{|c|}{}\\
\hline
&&&&&&&&\\
$N$ & \multicolumn{2}{c|}{$\beta=17.43$} &\multicolumn{2}{c|}{$\beta=17.63$}&\multicolumn{2}{c|}{$\beta=18.04$}&\multicolumn{2}{c|}{$\beta=18.375$}\\
&&&&&&&&\\
\hline
\multirow{18}{*}{$\,\,$ 5 $\,$ }
& 2 &  0.987907941(101)  &  2 &  0.988913542(92)  &  3 &  0.995941861(37)  &  3 &  0.996389536(40) \\
& 3 &  0.994581150(77)  &  4 &  0.997002160(59)  &  6 &  0.998733284(26)  &  6 &  0.998918380(28) \\
& 4 &  0.996596153(67)  &  6 &  0.998375494(49)  &  9 &  0.999309092(22)  &  9 &  0.999429207(23) \\
& 5 &  0.997553577(61)  &  8 &  0.998909560(44)  &  12 &  0.999536632(20)  &  12 &  0.999627702(21) \\
& 6 &  0.998103780(56)  &  10 &  0.999183460(40)  &  15 &  0.999653772(19)  &  15 &  0.999728338(19) \\
& 7 &  0.998456167(53)  &  12 &  0.999347001(38)  &  18 &  0.999723770(18)  &  18 &  0.999787622(18) \\
& 8 &  0.998698847(51)  &  14 &  0.999454648(36)  &  21 &  0.999769823(17)  &  21 &  0.999826101(18) \\
& 9 &  0.998875032(48)  &  16 &  0.999530489(34)  &  24 &  0.999802234(17)  &  24 &  0.999852832(17) \\
& 10 &  0.999008197(47)  &  18 &  0.999586668(33)  &  27 &  0.999826208(16)  &  27 &  0.999872362(16) \\
& 11 &  0.999112096(45)  &  20 &  0.999629914(32)  &  30 &  0.999844635(16)  &  30 &  0.999887194(16) \\
& 12 &  0.999195272(44)  &  22 &  0.999664233(31)  &  33 &  0.999859233(16)  &  33 &  0.999898810(15) \\
& 13 &  0.999263286(43)  &  24 &  0.999692146(30)  &  36 &  0.999871087(15)  &  36 &  0.999908138(15) \\
& 14 &  0.999319901(42)  &  26 &  0.999715315(29)  &  39 &  0.999880909(15)  &  39 &  0.999915786(15) \\
& 15 &  0.999367748(41)  &  28 &  0.999734877(28)  &  42 &  0.999889188(15)  &  42 &  0.999922165(14) \\
& 16 &  0.999408715(40)  &     &             &     &   &   45 &  0.999927567(14) \\
& 17 &  0.999444192(39)  &     &             &     &   &   48 &  0.999932199(14) \\
& 18 &  0.999475222(38)  &     &             &     &   &   51 &  0.999936216(14) \\
& 19 &  0.999502601(37)  &     &             &     &   &   54 &  0.999939734(14) \\
& 20 &  0.999526948(37)  &     &             &     &   &   57 &  0.999942842(13) \\
& 21 &  0.999548752(36)  &&&&&&\\
\hline
\end{tabular}
\begin{tabular}{|c|cc|cc|cc|cc|}
\hline
\multicolumn{9}{|c|}{}\\
\multicolumn{9}{|c|}{{\small$n_\cool\quad$ versus $P$}}\\
\multicolumn{9}{|c|}{}\\
\hline
&&&&&&&&\\
$N$ & \multicolumn{2}{c|}{$\beta=25.32$} &\multicolumn{2}{c|}{$\beta=25.70$}&\multicolumn{2}{c|}{$\beta=26.22$}&\multicolumn{2}{c|}{$\beta=26.65$}\\
&&&&&&&&\\
\hline
\multirow{18}{*}{$\,\,$ 6 $\,$ }
& 2 &  0.987682731(91)  &  2 &  0.988992297(83)  &  3 &  0.995895295(44)  &  3 &  0.996301620(38) \\
& 3 &  0.994488661(69)  &  4 &  0.997066786(54)  &  6 &  0.998728119(30)  &  6 &  0.998894654(25) \\
& 4 &  0.996548482(60)  &  6 &  0.998428346(45)  &  9 &  0.999309254(25)  &  9 &  0.999416738(21) \\
& 5 &  0.997524970(55)  &  8 &  0.998954355(41)  &  12 &  0.999538303(23)  &  12 &  0.999619462(19) \\
& 6 &  0.998085079(51)  &  10 &  0.999222771(38)  &  15 &  0.999655934(21)  &  15 &  0.999722171(17) \\
& 7 &  0.998443424(48)  &  12 &  0.999382292(36)  &  18 &  0.999726053(20)  &  18 &  0.999782619(16) \\
& 8 &  0.998690012(46)  &  14 &  0.999486826(34)  &  21 &  0.999772066(19)  &  21 &  0.999821810(16) \\
& 9 &  0.998868907(44)  &  16 &  0.999560155(33)  &  24 &  0.999804366(18)  &  24 &  0.999848998(15) \\
& 10 &  0.999004031(43)  &  18 &  0.999614247(31)  &  27 &  0.999828194(18)  &  27 &  0.999868828(15) \\
& 11 &  0.999109386(41)  &  20 &  0.999655715(30)  &  30 &  0.999846453(18)  &  30 &  0.999883861(14) \\
& 12 &  0.999193673(40)  &  22 &  0.999688489(29)  &  33 &  0.999860873(17)  &  33 &  0.999895608(14) \\
& 13 &  0.999262551(39)  &  24 &  0.999715042(29)  &  36 &  0.999872545(17)  &  36 &  0.999905019(14) \\
& 14 &  0.999319847(38)  &  26 &  0.999736998(28)  &  39 &  0.999882185(16)  &  39 &  0.999912714(14) \\
& 15 &  0.999368239(37)  &  28 &  0.999755466(27)  &  42 &  0.999890284(16)  &  42 &  0.999919113(13) \\
&  16 &  0.999409645(36)  &     &             &     &   &   45 &  0.999924517(13) \\
&  17 &  0.999445480(36)  &     &             &     &   &   48 &  0.999929133(13) \\
&  18 &  0.999476800(35)  &     &             &     &   &   51 &  0.999933122(13) \\
&  19 &  0.999504418(34)  &     &             &     &   &   54 &  0.999936600(13) \\
&  20 &  0.999528962(34)  &     &             &     &   &   57 &  0.999939660(12) \\
\hline
\end{tabular}
\caption{Each column reports the average plaquette $P$ as a function of $n_\cool$ for a given $\beta$ and $N$.}
\label{tab:rawdata_ene}
\end{table}

\end{document}